\documentclass[a4paper,11pt]{article}
\pdfoutput=1 
\usepackage{jheppub} 
\usepackage[T1]{fontenc} 

\usepackage{amsmath,amssymb}
\usepackage{amsfonts}
\usepackage{bm,bbm}
\usepackage{graphicx}
\usepackage{mathrsfs}
\usepackage{slashed}
\usepackage{booktabs}
\usepackage{tabu}
\usepackage[dvipsnames]{xcolor}
\usepackage[normalem]{ulem}
\usepackage{tikz}
\usepackage{soul}
\usepackage{extarrows}

\usepackage{hyperref}
\hypersetup{colorlinks,citecolor= blue,linkcolor= blue, urlcolor=blue}

\usepackage{xcolor}
\usepackage{ulem}
\usepackage{array}
\usepackage{verbatim}
\usepackage{epsfig}
\usepackage{multirow}
\usepackage{physics}
\usepackage{mathdots}
\usepackage{yhmath}
\usepackage{cancel}
\usepackage{siunitx}
\usepackage{amssymb}
\usepackage{gensymb}
\usepackage{tabularx}
\usepackage{extarrows}
\usepackage{booktabs}
\usetikzlibrary{fadings}
\usetikzlibrary{patterns}
\usetikzlibrary{shadows.blur}
\usetikzlibrary{shapes}

\newcommand{\be}{\begin{equation}}
\newcommand{\ee}{\end{equation}}
\newcommand{\ba}{\begin{array}}
\newcommand{\ea}{\end{array}}
\newcommand{\bea}{\begin{eqnarray}}
\newcommand{\eea}{\end{eqnarray}}

\definecolor{blue-violet}{rgb}{0.54, 0.17, 0.89}
\definecolor{amethyst}{rgb}{0.6, 0.4, 0.8}

\usepackage{ulem,fancyvrb}
\usepackage{xcolor}

\title{Probing ultralight isospin-violating mediators at GW170817}

\author[a]{Zuowei Liu}
\author[a]{and Zi-Wei Tang}

\affiliation[a]{Department of Physics, Nanjing University, Nanjing 210093, China}

\emailAdd{zuoweiliu@nju.edu.cn}
\emailAdd{ziweitang@smail.nju.edu.cn}

\abstract{
Gravitational wave (GW) signals arising from binary neutron star mergers offer new, sensitive probes to ultralight mediators. Here we analyze the GW signals in the GW170817 event detected by the LIGO/Virgo collaboration to impose constraints on the ultralight isospin-violating mediator that has different couplings to protons and neutrons. Neutron stars, which primarily consist of neutrons, are the ideal places to probe the isospin-violating mediator. Such a mediator can significantly alter the dynamics of the binary neutron star mergers, through both the long-range Yukawa force and the new dipole radiation. We compute the gravitational waveform by taking into account the new physics effects due to the isospin-violating mediator and use the Bayesian inference to analyze the gravitational wave data in the GW170817 event. We find that although the current fifth force experiments (including MICROSCOPE and EW) often provide more stringent constraints than the GW170817 data, in the parameter space where the isospin-violating force is completely screened by the Earth (namely, the Earth is charge neutral under this force), the GW170817 data offer the leading constraints: the upper bound on the neutron coupling is $f_n \lesssim 10^{-19}$ in the mediator mass range of $\simeq(3\times10^{-16},\,5\times10^{-14})$ eV. 
}

\begin{document} 
\maketitle
\flushbottom

\section{Introduction}

Ultralight bosons appear in a number of well-motivated 
new physics models beyond the standard model
\cite{Marsh:2015xka,Fabbrichesi:2020wbt,Jaeckel:2010ni,Essig:2013lka}. 
These particles are especially fascinating due to their potential to
mediate a new long-range force 
which is often referred to as a fifth force. 
Extensive experimental efforts have been devoted to searching 
for the effects of the fifth force across a wide range of distances and couplings 
\cite{Will:2014kxa}. 
One notable aspect of the fifth force searches is the exploration of the 
composition-dependent interactions \cite{Fischbach:2020kwt}, 
which could
lead to violations of the weak equivalence principle (WEP) 
\cite{Tino:2020nla, Adelberger:2006dh,Schlamminger:2007ht,
Wagner:2012ui,Touboul:2017grn, 
MICROSCOPE:2019jix,MICROSCOPE:2022doy,
Berge:2017ovy,Berge:2023sqt,Fayet:2017pdp,
Fayet:2018cjy,Talmadge:1988qz,Williams:2004qba,
Turyshev:2006gm,Williams:2012nc,Hofmann:2018myc}. 
The isospin-violating mediators, 
which have different couplings to neutrons and protons, 
are an intriguing type of composition-dependent interaction 
in fifth-force experiments 
and also of great interest in dark matter direct detection   
\cite{Kurylov:2003ra,Giuliani:2005my,Chang:2010yk,Feng:2011vu}. 
Moreover, 
ultralight isospin-violating mediators can also be searched for 
through gravitational wave signals in binary neutron star (BNS) mergers.

Thus, we use gravitational wave (GW) data from 
the BNS merger 
event GW170817 \cite{LIGOScientific:2017vwq} 
to probe the isospin-violating force.
The GW170817 event, 
the first BNS merger event detected by 
the LIGO/Virgo collaboration
\cite{LIGOScientific:2017vwq},
has a signal-to-noise ratio (SNR) of $32.4$, 
making it an ideal place to study various topics, 
including the properties of the NSs 
\cite{LIGOScientific:2018hze},
the equation-of-state (EoS) of nuclear matter 
\cite{LIGOScientific:2018cki},
and predictions from general
relativity (GR) and beyond 
\cite{LIGOScientific:2018dkp}.
The GW170817 event was detected by the two LIGO detectors
(Hanford and Livingston)
and the Virgo detector
(Cascina) \cite{LIGOScientific:2017vwq}. 
In contrast, the other BNS merger event, GW190425,
was only detected in the LIGO Livingston detector, 
with a lower SNR of $12.9$ \cite{LIGOScientific:2020aai}.
Moreover, optical counterparts of the GW170817 event \cite{DES:2017kbs,Cantiello:2018ffy} 
and the gamma-ray 
burst event, GRB 170817A, 
\cite{LIGOScientific:2017zic,LIGOScientific:2017ync} 
have been detected. 
The electromagnetic counterparts 
are instrumental in locating the host
galaxy of the BNS system,
facilitating the parameter estimation
of GW signal \cite{LIGOScientific:2018hze}.
Thus, we focus on the GW170817 event in our current analysis.

The ultralight mediator has two major effects on 
the dynamics of the BNS: 
First, it leads to a new long-range force 
so that the orbital frequency of the BNS 
can be modified significantly. 
Second, it provides a new radiation channel 
through which the BNS system can lose energy, 
if the mediator mass is small compared to the 
orbital frequency. 
This then leads to modifications to 
the GW signals arising from the BNS system; 
see e.g., Refs.~\cite{Croon:2017zcu,Alexander:2018qzg,Choi:2018axi,Fabbrichesi:2019ema,Dror:2019uea,Xu:2020qek,Zhang:2021mks,Huang:2018pbu,Hook:2017psm,Sagunski:2017nzb,Poddar:2023pfj} 
for previous studies. 
Orbital decays due to the radiation of ultralight mediators 
also have important effects in pulsar inspiraling 
\cite{Krause:1994ar,KumarPoddar:2019ceq,KumarPoddar:2019jxe,Dror:2019uea,Poddar:2023pfj,Seymour:2020yle,Seymour:2019tir}.
To our knowledge, the effects of ultralight
isospin-violating mediators on the GW170817 data have not been studied before. 
To probe the isospin-violating force,
we perform Bayesian analysis on
the GW170817 data 
\cite{LIGOScientific:2019lzm}, by
using the PyCBC inference 
\cite{Biwer:2018osg,alex_nitz_2023_7746324}. 
We find that 
the isospin-violating mediator with mass $\lesssim 10^{-11}$ eV 
can be probed by the GW170817 event,   
which consists of data with frequency up to $\sim$2000 Hz 
\cite{LIGOScientific:2017vwq}.

The parameter space of the isospin-violating mediators with 
mass $\lesssim 10^{-11}$ eV is 
constrained by a number of
experiments, including WEP, 
motions of asteroids 
\cite{Tsai:2021irw,Tsai:2023zza}
and planets
\cite{Talmadge:1988qz,KumarPoddar:2020kdz},
and black hole superradiance (BHSR) \cite{Baryakhtar:2017ngi}.
The dominant WEP experiments include the 
E\"ot-Wash (EW) torsion balance experiment 
\cite{Schlamminger:2007ht,Wagner:2012ui},
the lunar laser ranging (LLR) experiments 
\cite{Williams:2004qba,Turyshev:2006gm,Williams:2012nc,Hofmann:2018myc}, 
and the MICROSCOPE experiment
\cite{Touboul:2017grn,MICROSCOPE:2019jix,MICROSCOPE:2022doy,Berge:2017ovy,Berge:2023sqt}. 
Note that the signals in the WEP experiments depend on 
the charges of both the test mass and the attractor (gravity source). 
As emphasized in Ref.~\cite{Wagner:2012ui}, 
measurements with varied test masses and/or attractors are necessary, 
as charges may vanish for certain test masses or attractors.
For example, EW has used Be-Ti and Be-Al pairs as test masses \cite{Wagner:2012ui}, 
and MICROSCOPE has used Ti-Pt and Pt-Pt pairs \cite{MICROSCOPE:2022doy}. 
While substituting different test masses can be relatively straightforward, 
changing attractors often presents more of a challenge 
due to the limited number of nearby celestial bodies.
The most commonly used attractors include the Earth  
\cite{Schlamminger:2007ht,Wagner:2012ui,Touboul:2017grn,MICROSCOPE:2019jix,MICROSCOPE:2022doy,Berge:2017ovy,Berge:2023sqt} 
and the Sun \cite{Wagner:2012ui,Williams:2004qba,Turyshev:2006gm,Williams:2012nc,Hofmann:2018myc}.
Another critical factor in the fifth force experiments is 
the distance. 
For example,
experiments using the Sun as the attractor probe 
mediators with mass $\lesssim 10^{-18}$ eV, 
as the distance to the Sun is $\sim 1.5\times10^{8}$ km.
In contrast, experiments that use the Earth as the attractor, 
such as MICROSCOPE \cite{Touboul:2017grn,MICROSCOPE:2019jix,MICROSCOPE:2022doy,Berge:2017ovy,Berge:2023sqt}, 
can probe mediators in a larger mass range, 
$m \lesssim 3\times 10^{-14}$ eV, 
which is determined by the radius of the Earth. 
Consequently, the scenario 
where the charge of the Earth is zero and the mediator mass is in the range of
$\sim (10^{-18},3\times 10^{-14})$ eV  
cannot be probed by WEP experiments using the Sun as the attractor. 
It is thus of importance to employ alternative methods 
to explore this scenario.

Neutron stars, primarily composed of neutrons, 
differ significantly in composition from the Earth,
making BNS mergers an ideal place to probe the 
parameter space where the charge of the Earth is zero. 
Because the Earth consists of mainly heavy elements, 
its proton-to-neutron ratio is close to unity. 
In contrast, we find a proton-to-neutron ratio of $\sim (7-12)\%$ 
for NS with mass in the range of $\sim(1-2)\,M_\odot$, 
if the BSk24 EoS is used \cite{Pearson:2018tkr}. 
Moreover, 
the separation of the two neutron stars in the GW170817 event 
is $\sim(20, 400)$ km, which is smaller than the Earth radius 
by one to two orders of magnitude, 
so that it has the potential to probe 
mediators with an even larger mass range of  
$m\lesssim 10^{-11}$ eV. 
In the mass range of $\sim(10^{-16}-10^{-13})$ eV, 
the dominant WEP constraints come from the WEP 
experiments that use Earth as the attractor, 
such as the MICROSCOPE experiment 
\cite{Touboul:2017grn,MICROSCOPE:2019jix,MICROSCOPE:2022doy,Berge:2017ovy,Berge:2023sqt}, 
and the EW experiment \cite{Schlamminger:2007ht,Wagner:2012ui}.  
We find that the GW170817 data can offer leading constraints 
for this mass range, if the Earth charge vanishes.

To correctly interpret the GW170817 constraints, 
as well as other experiments that use celestial bodies 
as the source of the external field, 
it is important to properly take into account the effects 
due to the finite sizes of these celestial bodies 
when the Compton wavelength of the mediator 
becomes small compared to the size of the celestial bodies 
\cite{Adelberger:2003zx}. 
To this end one needs to integrate over the charge 
density distribution within the celestial bodies. 
In our analysis, we use 
the Tolman-Oppenheimer-Volkoff (TOV) equation 
\cite{Tolman:1939jz,Oppenheimer:1939ne}
together with an EoS
\cite{Pearson:2018tkr} 
to determine 
the proton and neutron density distributions 
for NSs with different masses, 
and then use these distributions 
to compute the charge of the NSs for the isospin-violating force. 
We also provide a simple analytic expression of the Earth 
charge, correcting a previous erroneous formula in 
Ref.~\cite{Berge:2017ovy}.

The rest of the paper is organized as follows. 
In section \ref{sec:2} we discuss the effects of the 
ultralight isospin-violating mediator on the BNS dynamics, 
and provide calculations of the NS charge 
by taking into account the charge density 
distributions in the NS. 
In section \ref{sec:waveform} we compute the GW waveform
in the presence of the fifth force, 
and then use it to perform Bayesian analysis of the GW170817 data in section \ref{sec:para}. 
We discuss various fifth force constraints in 
section \ref{sec:other:constraints}, 
including 
constraints from asteroids and planets data, 
from WEP experiments, 
and from BHSR.
We present the GW170817 constraints on 
the ultralight isospin-violating mediators 
along with 
other experimental constraints in 
section \ref{sec:results}, 
where we also discuss two interesting cases: 
the baryon number case, 
and the case where the Earth charge is zero. 
We summarize our findings in section \ref{sec:suma}. 
The detailed calculations on the neutron star 
properties are given in Appendix 
\ref{sec:app-nucleon-fraction}.

\section{Ultralight isospin-violating mediator and BNS}
\label{sec:2}

In this study, we utilize the
data from the GW170817 gravitational
wave event \cite{LIGOScientific:2017vwq}
to constrain the isospin-violating force. 
We consider the following interaction Lagrangian 
\begin{equation}\label{eq:lagrangian}
    \mathcal{L}_{\rm int} \supset 
    V_\mu (f_n \bar{n}\gamma^\mu n 
    + f_p \bar{p}\gamma^\mu p),
\end{equation}
where $V_\mu$ is the vector field 
that couples to proton $p$ and neutron $n$
with couplings $f_p$ and $f_n$, respectively. 
If $V_\mu$ is an ultralight particle, 
it can affect the BNS dynamics in two different ways: 
First, its long-range Yukawa force can 
alter the orbital frequency of the BNS significantly. 
Second, 
in addition to the gravitational radiation, 
there is a new dipole radiation channel
through which the BNS can lose energy.

\subsection{Yukawa force and orbiting frequency}
\label{sec:yukforce}

If the vector field $V_\mu$ has a sufficiently small mass, 
it generates a new long-range force. 
If the mass $m_V \lesssim 10^{-11}$ eV, 
the long-range force has a characteristic scale 
$\lambda = 1/m_V \gtrsim 20$ km, 
thus having the potential to significantly
alter the orbiting speed of the binary neutron stars, 
which is primarily governed by gravity.\footnote{The 
separation $r$ of the NSs is related to the GW 
frequency $f$ via $r^3=G(m_1+m_2)/(\pi f)^2$. 
To analyze the GW170817 data,  
Ref.~\cite{LIGOScientific:2018hze} used 
$23 \leq f \leq 2048$ Hz, thus corresponding to the range of 
$21 \lesssim r \lesssim 410$ km.}
The new long-range force can increase (decrease) 
the orbiting speed of the BNS, 
if it is attractive (repulsive) 
between the two neutron stars. 
At the leading order, 
the total force acting on the two neutron stars is the sum of 
the Newtonian gravity and the Yukawa-type force 
mediated by the $V_\mu$ field: 
\cite{Croon:2017zcu,Alexander:2018qzg, Dror:2019uea}
\begin{equation}\label{eq:force}
F(r) = \frac{Gm_1m_2}{r^2}
\left[1-\alpha\,
e^{-m_Vr}\left(1+m_Vr\right)\right],
\end{equation}
where 
$G$ is the gravitational constant, 
$m_1$ and $m_2$ are the masses of the neutron stars, 
$r$ is the distance between the two neutron stars, 
$m_V$ is the mass of $V_\mu$, 
and $\alpha$ is the parameter 
that characterizes the
relative strength of the new Yukawa 
force induced by $V_\mu$. 
The parameter $\alpha$ is given by
\begin{equation}\label{eq:alpha}
    \alpha\equiv\frac{Q_1 Q_2}{4\pi Gm_1m_2},
\end{equation}
where $Q_1$ and $Q_2$ are the charges of the two neutron stars.
A positive (negative) $\alpha$ 
indicates a repulsive (attractive) Yukawa force.

In the presence of such a Yukawa-type force, 
the orbital frequency of the BNS
system is given by the modified Kepler's law 
\cite{Croon:2017zcu,Alexander:2018qzg, Dror:2019uea}:
\begin{equation}\label{eq:Kepler}
    \omega^2=\frac{GM}{r^3}\left[1-\alpha\left(1+m_Vr\right)e^{-m_Vr}\right]
\end{equation}   
where $M=m_1+m_2$.
The $\alpha$ term resulting from the Yukawa force 
exhibits a distinct radial dependence 
as compared to the gravity term, 
which is crucial 
in distinguishing between the Yukawa and gravity forces 
in the waveform analysis of the gravitational wave data.

\subsection{Dipole radiation}

If the orbital frequency of the BNS system is larger than
the mass of the light mediator (in natural units), 
the BNS system can lose energy by radiating the 
$V_\mu$ field, 
thus presenting a new energy-loss channel 
in addition to the gravitational radiation channel. 
Analogous to the electrodynamics, 
the radiation due to the massive 
$V_\mu$ field can be expressed as a set of multipole
expansion terms \cite{Krause:1994ar}, 
with the leading order contribution 
from the electric dipole radiation\footnote{Next leading 
order contributions are the magnetic dipole and the electric
quadrupole radiation \cite{Krause:1994ar}.}
The radiation power (energy emitted per unit time)
in the electric dipole radiation is given by 
\cite{Krause:1994ar,
Croon:2017zcu,
Alexander:2018qzg,
KumarPoddar:2019ceq},
\begin{equation}\label{eq:diopleradiation}
    P_{V} = \frac{2}{3}G\gamma\mu^2\omega^4r^2
    \left[1+\frac{1}{2}\left(\frac{m_V}{\omega}\right)^2\right]
    \sqrt{1-\left(\frac{m_V}{\omega}\right)^2},
\end{equation}
where 
$\omega$ is the orbital frequency of the BNS system,
$m_V$ is the mediator mass,
$\mu\equiv m_1m_2/M$ is the reduced mass of the BNS system, 
and $\gamma$ is given by 
\begin{equation}\label{eq:gamma}
    \gamma\equiv\frac{1}{4\pi G}\left(\frac{Q_1}{m_1}-
    \frac{Q_2}{m_2}\right)^2.
\end{equation}
Note that we have used a zero orbital eccentricity 
to obtain Eq.~\eqref{eq:diopleradiation}, 
which is justified since the effect of 
the orbital eccentricity in GW170817 
is negligible \cite{Lenon:2020oza}.

To compute the orbiting frequency of the 
BNS system, we use the energy conservation  
to write 
\begin{equation}\label{eq:energyconservation}
    \frac{d}{dt}\left(E_{G}+E_V\right) = -\left(P_{G}+P_V\right),
\end{equation}
where $E_{G}$ is the total energy of the BNS system
that is predicted by GR, $P_{G}$ is the gravitational
radiation power, 
and $E_V$ is the Yukawa correction
to the potential energy of the binary system:
\begin{equation}\label{eq:yukawapotential}
    E_V = 
    \frac{Gm_1m_2}{r}\alpha
    e^{-m_Vr}.
\end{equation}
In our analysis we use 
the binary energy at the Newtonian level \cite{Dror:2019uea} 
\begin{equation}
\label{eq:GRenergy}
E_{G} = \frac{1}{2}\mu r^2 \omega^2-\frac{Gm_1m_2}{r},
\end{equation}
and the 
quadrupole gravitational radiation power
\cite{carroll_2019,Alexander:2018qzg}
\begin{equation}
\label{eq:GWradiation}
P_{G}=\frac{32}{5}G\mu^2r^4\omega^6.
\end{equation}

\subsection{NS Charge}
\label{sec:charge:NS}

Both the parameter $\alpha$, which enters the new Yukawa force, 
and the parameter $\gamma$, which enters the new radiation power,   
depend on the charge $Q$ of the NS. 
When the wavelength of the mediator is sufficiently large so that 
the NS can be treated as a point particle, 
the charge $Q$ of the NS can be obtained via a simple algebraic summation. 
However, when the size of the NS is comparable to 
or larger 
than the wavelength of the mediator,  
one has to properly take into account the effects 
due to the finite size of the NS and 
the charge distribution within the NS.

For a spherically symmetric object with a uniform density, 
the charge is given by \cite{Adelberger:2003zx}
\begin{equation}\label{eq:Quniform}
Q_{u} = Q_{\rm pt} \, 
\Phi\left(\frac{R}{\lambda}\right)
\end{equation}
where 
$Q_{\rm pt}$ is the charge 
when the object can be treated as a point particle, 
$\Phi(x)\equiv3(x\cosh x-\sinh x)/x^3$, 
$R$ is the radius of the object, 
and $\lambda$ is the Compton
wavelength of the mediator. 
\footnote{Note that $\Phi(x) \to 1$, when $x \to 0$.} 
It is thus straightforward to obtain the charge of 
a spherically symmetric object 
with a charge density distribution: 
\begin{equation}\label{eq:Qint}
Q = 4 \pi \lambda \int_0^{R} dr \, r \, q(r) \, 
\sinh \left(\frac{r}{\lambda} \right), 
\end{equation}
where 
$r$ is the radial coordinate, and 
$q(r)$ is the charge density distribution. 
Note that in the case of constant $q$, 
one has $Q_{\rm pt} = q (4\pi R^3/3)$.

For the isospin-violating mediator, 
the charge density distribution $q(r)$ in Eq.~\eqref{eq:Qint}
is given by 
\begin{equation}
    q(r)=f_p n_p(r)+f_n [n(r)-n_p(r)],
\end{equation}
where $n_p$ ($n$) is the proton (nucleon) number density. 
In the long-wavelength limit, 
the charge of the NS becomes 
\begin{equation}\label{eq:Qpt}
    Q_{\rm pt} =f_p Z+f_n N,
\end{equation}
where $Z$ ($N$) is the proton (neutron) number of the NS. 
To obtain the charge density distribution, 
we first determine the total nucleon number density $n(r)$, 
by solving the TOV equation \cite{Tolman:1939jz,Oppenheimer:1939ne} 
together with an EoS.
We then use the proton-to-nucleon fraction $Y_p(n)$, 
to obtain the proton and neutron distributions.
See Appendix \ref{sec:app-nucleon-fraction} for the detailed analysis.

We note that throughout our analysis, 
we will mainly use BSk24 \cite{Pearson:2018tkr} 
(the default EoS unless explicitly specified) for the EoS. 
For comparison, we also consider an alternative EoS, 
BSk22 \cite{Pearson:2018tkr}, 
in Figs.~(\ref{Fig:couplings-ratio},\ref{fig:nrYpn},\ref{fig:rhocfraction}).

\subsection{GW170817}

In our analysis we focus on the GW170817 event, 
where the primary NS has a mass of  
$m_1=1.46^{+0.12}_{-0.10}\,M_\odot$
and a radius of  
$11.9^{+1.4}_{-1.4}$ km, 
and the secondary NS has a mass of 
$m_2=1.27^{+0.09}_{-0.09}\,M_\odot$
and a radius of 
$11.9^{+1.4}_{-1.4}$ km
\cite{LIGOScientific:2018mvr,LIGOScientific:2018cki}. 
For a mediator with mass of 
$m_V \gtrsim 10^{-11}$ eV 
($\lambda \lesssim 20$ km), 
\footnote{In our analysis we use natural units. 
The conversion between eV and meter is
$1$ eV$^{-1}$ $=$ $1.9733\times 10^{-7}$ m.}
one has to take into account the effects of the charge 
distribution of the NS. 
The total proton and neutron numbers for the two NSs are: 
$N_1 \simeq 1.316\times 10^{56}$, 
$Z_1 \simeq 1.536\times 10^{57}$, 
$N_2 \simeq 1.052\times 10^{56}$, and 
$Z_2 \simeq 1.359\times 10^{57}$.

\begin{figure}[htbp]
\begin{centering}
\includegraphics[width=0.45 \textwidth]{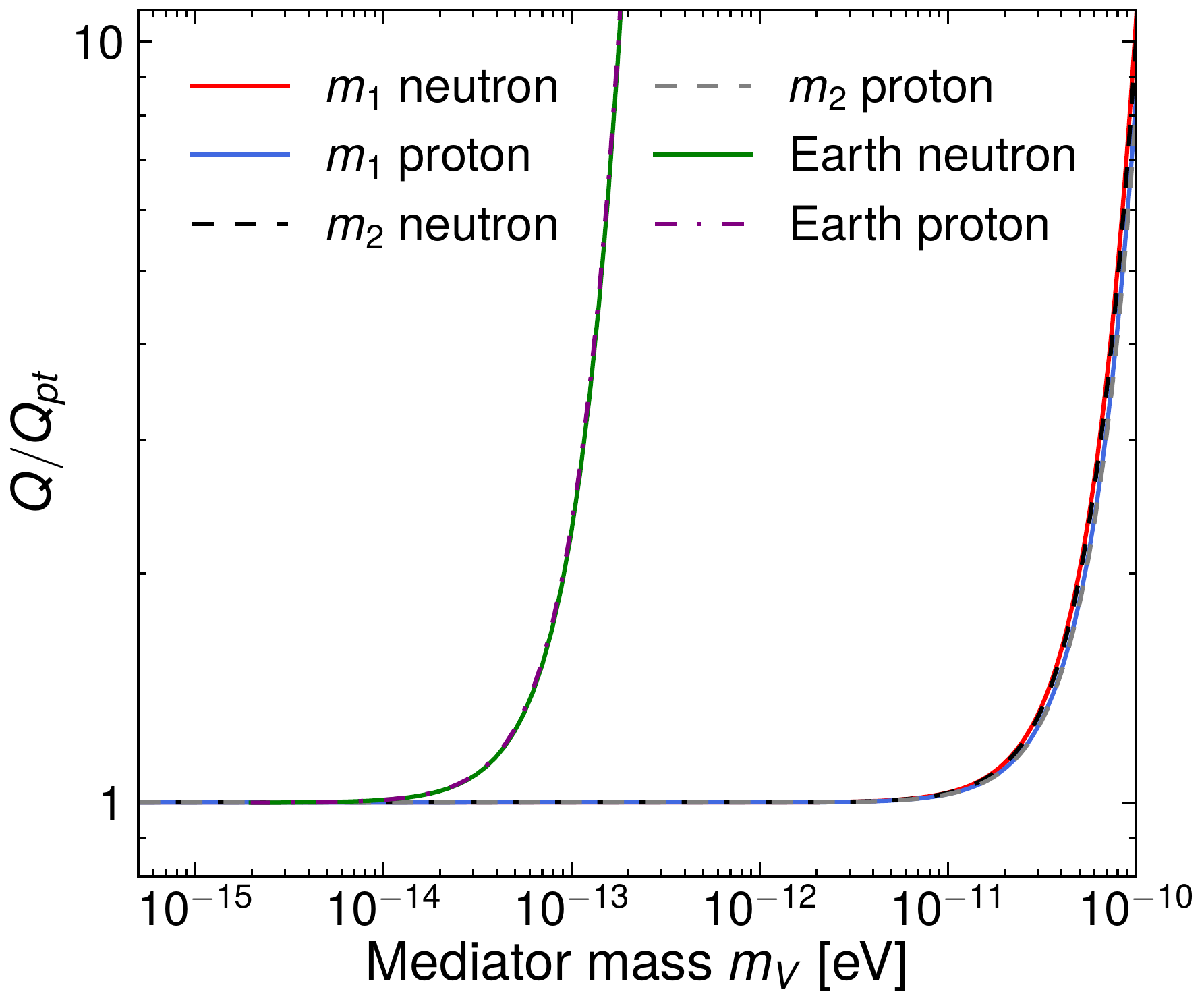}
\includegraphics[width=0.45 \textwidth]{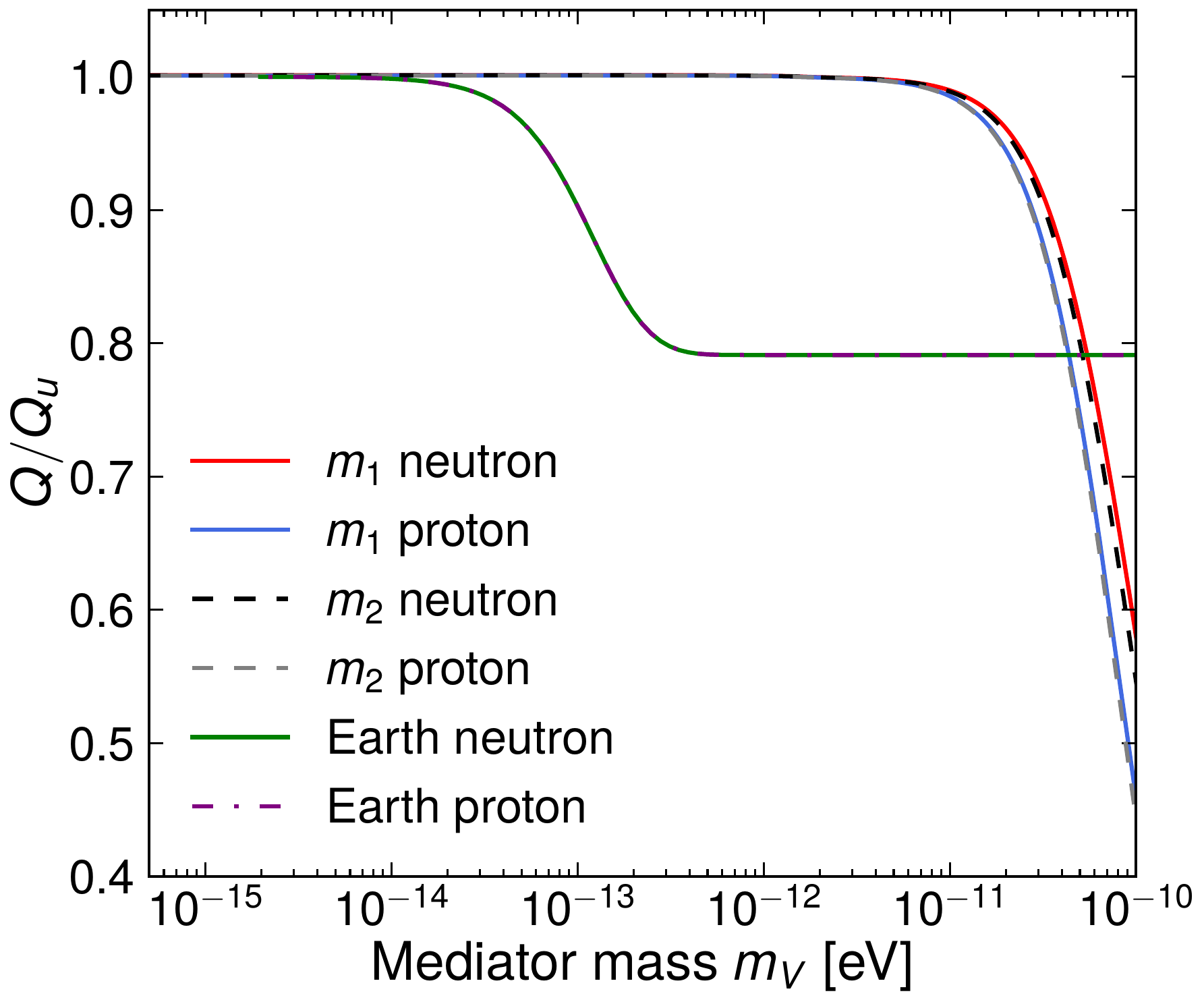}
\caption{The proton and neutron charges 
of the two NSs (denoted as $m_1$ and $m_2$) in 
the GW170817 event 
and of the Earth as a function of the mediator mass. 
We normalize the charge to  
the long-wavelength limit $Q_{\rm pt}$ ({\bf left panel}), 
and to $Q_u=Q_{\rm pt} \Phi(R/\lambda)$ 
which is given in Eq.~\eqref{eq:Quniform} ({\bf right panel}). 
The masses of the two NSs in GW170817 are $m_1=1.46\,M_\odot$
and $m_2=1.27\,M_\odot$. 
}
\label{fig:Qmv}
\end{centering}
\end{figure}

Fig.~\ref{fig:Qmv} shows the proton and neutron charges 
of the two NSs in GW170817 as a function of the mediator
mass. 
In the left panel figure of Fig.~\ref{fig:Qmv}, 
the charges are normalized with respect to the 
long-wavelength limit $Q_{\rm pt}$; 
the ratio $Q/Q_{\rm pt}$ starts to creases from unity 
where the mediator mass becomes $\gtrsim 10^{-11}$ eV. 
In the right panel figure of Fig.~\ref{fig:Qmv}, 
the charges are normalized with respect to 
$Q_u=Q_{\rm pt} \Phi(R/\lambda)$ 
which is given in Eq.~\eqref{eq:Quniform}; 
the ratio $Q/Q_u$ is significant when 
$m_V \gtrsim 10^{-11}$ eV, 
which shows that the integral over the nucleon density distribution 
provides a substantial contribution.

Note that 
the isospin-violating mediator with mass $\lesssim 10^{-11}$ eV is  
also constrained by other experiments that use Earth or Sun as the target. 
To correctly interpret the limits from these experiments, 
one has to compute the charge of the Earth/Sun 
by taking into account the charge distributions within these objects. 
We discuss these constraints in section \ref{sec:other:constraints}.

\section{Inspiral waveform}
\label{sec:waveform}

The gravitational waveform in the frequency domain 
(the Fourier representation
of the GW strain) can be written as
\cite{Yunes:2009yz,Buonanno:2009zt}
\begin{equation}
    h_i(f)=\mathcal{A}_i(f)e^{-i\Psi(f)},
\end{equation}
where 
$f$ is the GW frequency, 
$i=(+,\times)$ denote 
the ``plus'' and ``cross'' polarizations of the GW, 
$\mathcal{A}_i(f)$ is the amplitude, 
and $\Psi(f)$ is the phase. 
In our analysis, 
we compute the amplitudes and the phase via  
\begin{align}
    \label{eq:decompsedamp} \mathcal{A}_i&=\mathcal{A}_{i,G}+\mathcal{A}_{i,V},\\
    \label{eq:decompsedpha} \Psi&=\Psi_G+\Psi_V,
\end{align}
where $\mathcal{A}_{i,G}$ and $\Psi_G$ are the
amplitudes and phase in the TaylorF2 model
in the LALsuite \cite{lalsuite},
$\mathcal{A}_{i,V}$ and $\Psi_V$ are the
contributions from the new vector mediator.

\subsection{Amplitude}

In the stationary phase approximation (SPA), 
the waveform amplitude is given by 
\cite{Yunes:2009yz,Alexander:2018qzg}
\begin{align}
    \label{eq:ampplus} &
    \mathcal{A}_+(f) =  -\frac{1+\cos^2 \theta_{JN} }{2}\frac{2G\eta M}{D_L}
    \left(\pi f\right)^2r^2(t_s)\sqrt{\frac{\pi}{\dot{\omega}(t_s)}},\\
    \label{eq:ampcross} &
    \mathcal{A}_\times (f) = i\cos \theta_{JN} \frac{2G\eta M}{D_L}
    \left(\pi f\right)^2r^2(t_s)\sqrt{\frac{\pi}{\dot{\omega}(t_s)}},
\end{align}
where 
$\theta_{JN}$ is the inclination angle 
such that $\cos\theta_{JN}=\hat{\bf J} \cdot \hat{\bf N}$
with ${\bf J}$ being the angular momentum of the
BNS system and ${\bf N}$ being the line of sight, 
$\eta \equiv m_1m_2/M^2$, 
$D_L$ is the luminosity distance of the BNS system, 
$t_s$ is defined as the stationary point such that
$\omega(t_s)\equiv\dot{\phi}(t_s)=\pi f$ 
with $\phi$ being the orbital phase of the BNS system, 
and $r(t_s)$ and $\dot{\omega}(t_s)$ are the  
$r$ and $\dot{\omega}$ values at time $t_s$, 
respectively.
Because $\omega=\pi f$ at $t_s$, 
we determine $r(t_s)$ and $\dot{\omega}(t_s)$ by 
evaluating $r(\omega)$ and $\dot{\omega}(\omega)$ 
at $\omega=\pi f$. 
We determine $r(\omega)$ from Eq.~\eqref{eq:Kepler}. 
In our analysis we treat $\alpha$ as a small parameter 
and expand $r(\omega)$ in powers of $\alpha$ \cite{Alexander:2018qzg}: 
\begin{equation}
r(\omega) = r_0(\omega) + \alpha r_1(\omega) + \mathcal{O}(\alpha^2), 
\end{equation}
which is then inserted into Eq.~\eqref{eq:Kepler}. 
We thus find 
\begin{equation}\label{eq:separation}
    r(\omega)=\frac{GM}{v^2}\left[1-\frac{\alpha}{3}\left(1+
    \beta v^{-2}\right) \text{exp}\left(-\beta v^{-2}\right)\right]
    + \mathcal{O}(\alpha^2), 
\end{equation}
where $\beta\equiv GMm_V$ and
$v\equiv(GM\omega)^{1/3}$. 
We compute $\sqrt{1/\dot{\omega}(\omega)}$ via
\begin{equation}\label{eq:sqrtomegadot}  
    \sqrt{\frac{1}{\dot{\omega}}} = 
    \sqrt{\frac{dE/d\omega}{dE/dt}} = 
    \sqrt{-\frac{E'_G+E'_V}{P_G+P_V}},
\end{equation}
where 
$P_V$ is given 
Eq.~\eqref{eq:diopleradiation}, 
$E_V$ is given 
Eq.~\eqref{eq:yukawapotential},
$E_G$ is given 
Eq.~\eqref{eq:GRenergy}, 
and $P_G$ is given 
Eq.~\eqref{eq:GWradiation}. 
In the last step, we have used the energy conservation 
relation in Eq.~\eqref{eq:energyconservation}, 
and the prime denotes the derivative with 
respect to $\omega$.
We next expand $r^2\dot\omega^{-1/2}$ to find 
the leading terms in 
$\alpha$ and $\gamma$, 
which are then used in 
Eqs.~(\ref{eq:ampplus}-\ref{eq:ampcross})
to obtain $\mathcal{A}_{+,V}(f)$ 
and $\mathcal{A}_{\times,V} (f)$. 
We add these two NP terms to the TaylorF2 amplitudes.

\subsection{Phase}

In the SPA, the phase of the GW waveform is given by
\cite{Alexander:2018qzg,Yunes:2009yz}
\begin{equation}\label{eq:frequencyphase}
    \Psi(f) = 2\pi ft_s-2\phi(t_s)-\frac{\pi}{4}.
\end{equation}
We first compute $t(\omega)$ and $\phi(\omega)$ via 
\cite{Buonanno:2009zt}
\begin{align}
\label{eq:timeintegral}    
t(\omega)&=\int_{\omega_c}^\omega d\omega'
\frac{1}{\dot{\omega}'}+t_c, \\
\label{eq:phaseintegral}  
\phi(\omega)&=\int_{\omega_c}^\omega d\omega^\prime
\frac{\omega'}{\dot{\omega}^\prime}+\phi_c,
\end{align}
where 
$\omega_c$, $t_c$ and $\phi_c$
are 
the orbital frequency, 
time, 
and the orbital phase at
coalescence, respectively. 
Following the LALsuite's TaylorF2 routines \cite{lalsuite}, we 
identify the coalescence as the moment
that the orbital frequency approaches infinity, 
$\omega_c \rightarrow +\infty$.  
We compute $1/\dot{\omega}$ via 
\begin{equation}\label{eq:omegadot}  
    \frac{1}{\dot{\omega}} = 
    \frac{dE/d\omega}{dE/dt} = 
    -\frac{E'_G+E'_V}{P_G+P_V}.
\end{equation}
We determine $t_s$ and $\phi(t_s)$ 
by evaluating $t(\omega)$ and $\phi(\omega)$ at $\omega=\pi f$. 
To determine the NP contribution to the phase,  
we expand $1/\dot{\omega}$ to find 
the leading terms in $\alpha$ and $\gamma$, 
which are then inserted into  
Eqs.~(\ref{eq:timeintegral}-\ref{eq:phaseintegral}) 
and further in Eq.~\eqref{eq:frequencyphase} 
to obtain the NP phase $\Psi_V$.
We add $\Psi_V$
to the phase term in the TaylorF2 template.

We note that the TaylorF2 template is obtained 
at the 3.5 post-Newtonian (PN) order \cite{Buonanno:2009zt}. 
The PN order in Ref.~\cite{Buonanno:2009zt} refers to the 
power of $x=v^2$ beyond the leading order contributions, where 
$v=(GM\omega)^{1/3} = (GM\pi f)^{1/3}$ is the 
characteristic velocity of the binary system. 
For example, at the GW frequency $f=100$ Hz, 
where the LIGO detectors reach their maximum sensitivity \cite{KAGRA:2013rdx}, 
one has $v\simeq 0.16$ for the BNS system in the GW170817 event, 
and the 3.5PN order expression of $E_G$ deviates 
from the Newtonian expression only by $\sim 2\%$. 
To obtain new physics contributions, however, 
we have avoided using the 3.5PN expansion. 
This is because 
the new physics contributions at the 3.5 PN order are so complex 
that the phase term in
Eq.~\eqref{eq:frequencyphase} cannot be analytically obtained; 
a 
numerical treatment on the integrals in 
Eqs.~(\ref{eq:timeintegral}-\ref{eq:phaseintegral}) is not favored, 
as
they could produce 
significant
uncertainties or even 
hidden errors in our MCMC analysis. 
Therefore, 
following Refs.~\cite{Dror:2019uea,Zhang:2021mks}, 
we have used Newtonian level expressions  
to obtain the new physics contributions in this section. 
For example, the 
$E_G$ formula in Eq.~\eqref{eq:GRenergy} 
and the $P_G$ formula in Eq.~\eqref{eq:GWradiation} 
are derived at the Newtonian level 
(0PN) 
and then used in 
Eq.~\eqref{eq:sqrtomegadot} and 
Eq.~\eqref{eq:omegadot} 
to obtain the new physics contributions  
to the amplitude and to the phase.

To estimate the difference between our approach
and the one where a 3.5PN analysis can be reliably carried out, 
we compute the new physics contributions  
that are proportional to $\alpha$ in Eq.~\eqref{eq:omegadot}, 
by using $E_G$ and $P_G$ at both the 0PN and 3.5PN orders.  
We find that our approach underestimates
the new physics contributions 
by $\sim 10\%$ 
at $f \sim 100$ Hz for $m_V\lesssim 10^{-13}$ eV. 
Because of the dominance of the BHSR constraints for 
$5 \times 10^{-14}$ eV $\lesssim m_V \lesssim 2\times10^{-11}$ eV, 
our GW analysis only probes new parameter space outside 
this mass range; see section \ref{sec:other:constraints} for details.
Fig.~\ref{fig:results} shows that  
for $m_V \lesssim 5 \times 10^{-14}$ eV, 
the 3$\sigma$ upper bound on $\alpha$ is in the 
range of $(0.3-0.4)$. 
Therefore, the next-leading-order (NLO) corrections of new physics, 
which are of order $\alpha$, 
seem rather significant. 
Thus, in the parameter space of interest 
(namely $m_V \lesssim 5 \times 10^{-14}$ eV), 
the uncertainty due to the use of the 0PN expression 
is smaller than that of neglecting the NLO corrections of new physics, 
by a factor of $\sim(3-4)$.
We leave the analysis of new physics contributions with 
high-PN orders and/or high-$\alpha$ orders to a future study.

\section{Parameter estimation}
\label{sec:para}

To search for the isospin-violating force, 
we use the Markov-Chain Monte Carlo (MCMC) method 
to sample the parameter space.
We select 40 different mediator masses 
in the range of $[10^{-14},10^{-10}]$ eV, 
equally spaced in the log-scale, 
as shown in Fig.~\ref{fig:results}. 
\footnote{GW170817 constraints on $\alpha$ become large 
for the mediator mass $\gtrsim 10^{-10.5}$ eV 
so that perturbative analysis is no longer valid.
Thus, we do not show model points beyond $\gtrsim 10^{-10.5}$ eV 
in Fig.~\ref{fig:results}.}
For each mediator mass, 
we perform the Bayesian analysis of the 
GW170817 data \cite{LIGOScientific:2017vwq,LIGOScientific:2019lzm}, 
by using PyCBC \cite{Biwer:2018osg,alex_nitz_2023_7746324}.
We evaluate the posterior probability $p(\vec{\theta}|d,H)$, 
where 
$d$ is the GW170817 data, 
$H$ is the signal model, 
and 
$\vec{\theta}$ denotes the parameter 
space of the signal model $H$.   
We obtain the signal model $H$ by modifying 
the TaylorF2 template \cite{LIGOScientific:2018hze}
to take into account the following 
new physics effects due to the isospin-violating mediator: 
(1) the effect on the BNS inspiral dynamics 
due to the additional Yukawa type force, 
as given in Eq.~\eqref{eq:force}; 
(2) the new energy loss channel due to the radiation of the
light mediator, 
as given in Eq.~\eqref{eq:diopleradiation}.

In our analysis the parameter space $\vec{\theta}$ 
is spanned by 11 different parameters, 
which can be grouped into four 
categories: 

\begin{itemize}

\item 

Neutron star parameters: 

\begin{itemize}
\item 
the masses of the two neutron stars, $m_1$ and $m_2$, 
\item 
the aligned spins of the two neutron stars, 
$\chi_{1z}$ and $\chi_{2z}$, 
where the $z$ direction is along the orbital
angular momentum of the BNS system, 
\item 
and the tidal deformabilities of the two neutron stars, 
$\Lambda_1$ and $\Lambda_2$; 
\end{itemize}

\item 
BNS parameters: 

\begin{itemize}
\item 
the inclination angle $\theta_{\text{JN}}$, 
where $\cos\theta_{\text{JN}}=\hat{\bf{J}}\cdot\hat{\bf{N}}$
with $\bf{J}$ being the angular momentum of the
BNS system and $\bf{N}$ being the line of sight;

\item 
the polarization of BNS $\psi$, 
which is the angle between the 
natural polarization basis of the GW
and the reference polarization basis \cite{GWLects};

\end{itemize}

\item 
The coalescence time $t_c$ of the merger;

\item 
New physics parameters: 
\begin{itemize}
\item 
$\alpha$, the relative strength
of the Yukawa-force, 
as given in Eq.~\eqref{eq:alpha}; 
\item 
$\gamma$, the parameter for the new radiation channel, 
as given in Eq.~\eqref{eq:gamma}; 
\end{itemize}

\end{itemize}

In our analysis, we fix the following parameters 
for the GW170817 event: 
luminosity distance $D_\text{L}=40.7$ Mpc 
and the sky location (RA, Dec) = ($197.450374^\circ, -23.381495^\circ$) 
\cite{LIGOScientific:2018hze, DES:2017kbs, Cantiello:2018ffy}. 
Additionally, 
to better converge the stochastic samplers, 
we marginalize the coalescence phase $\phi_c$ in estimating the likelihood function
\cite{LIGOScientific:2018hze,Veitch:2014wba}. 
Table \ref{tab:prior} shows the 
priors of the 11 parameters that are being sampled.

\begin{table*}[htbp]
    \centering
    \renewcommand{\arraystretch}{1.1}
    \begin{tabular}{|c|c|c|c|}
    \hline
    Parameter & prior
    & Parameter & prior \\ \hline
    $m_1$, $m_2$ ($M_{\odot}$) & $\left(1.0, 2.0\right)$ 
    & $\chi_{1z}$, $\chi_{2z}$ & $\left(-0.05, 0.05\right)$
    \\ \hline
    $\Lambda_1$, $\Lambda_2$ & $\left(0, 5000\right)$ 
    & $\iota$ (rad) & $\left(0, 2\,\pi\right)$
    \\ \hline
    $\psi$ (rad) & $\left(0, 2\,\pi\right)$
    & $\Delta{t_c}$ (s) & $\left(-0.1, 0.1\right)$
    \\ \hline
    $\alpha$ & $\left(-0.1, 0.1\right)^*$
    & $\gamma$ & $\left(0, 0.1\right)^*$
    \\ \hline
\end{tabular}
\caption{The typical prior distributions
for the sampled parameters.
We use flat (uniform) distributions
for the parameters with boundaries
that are shown in the brackets. $\Delta{t_c}$
means we sample the coalescence time around the (LIGO) gps time
$1187008882.4$ with
a range $\Delta{t_c}$.
For certain sampler runs, we may
adjust the priors of the fifth force
parameters $\{\alpha,\,\gamma\}$ since
they vary significantly when changing
the value of the mediator mass.
}
\label{tab:prior}
\end{table*}

To compute the posterior probability 
$p(\vec{\theta}|d,H)$, we
analyze the GW170817 data
in the frequency range of
$f_{\text{low}}\leq f \leq f_{\text{ISCO}}$,
where $f_{\text{low}} = 20$ Hz \cite{Zhang:2021mks}, 
and $f_{\text{ISCO}}$ is the frequency at
the innermost stable circular orbit (ISCO)
of the BNS system \cite{Alexander:2018qzg}:
\begin{equation}
    f_{\text{ISCO}}=(4.4\times 10^3\, \text{Hz})
    \left(\frac{m_1+m_2}{M_\odot}\right)^{-1}\simeq 1.6\times 
    10^3\,\text{Hz},
\end{equation}
where in the last step,  
we have used $m_1=1.46\,M_\odot$
and $m_2=1.27\,M_\odot$ \cite{LIGOScientific:2018hze}.
The TaylorF2 template
no longer accurately describes the BNS dynamics 
when the frequency exceeds $f_{\text{ISCO}}$ 
\cite{LIGOScientific:2018hze}.

\section{Other experimental constraints}
\label{sec:other:constraints}

In this section we discuss other relevant experimental 
constraints on the isospin-violating mediators with 
mass $\lesssim 10^{-11}$ eV. 
These constraints can be categorized into three types: 
those arising from motions of planets
\cite{Talmadge:1988qz,KumarPoddar:2020kdz} and asteroids \cite{Tsai:2021irw,Tsai:2023zza}, 
experiments that test the WEP, 
and BHSR \cite{Baryakhtar:2017ngi}. 
The WEP experiments include the 
EW torsion balance experiment 
\cite{Schlamminger:2007ht,Wagner:2012ui},
the LLR experiments 
\cite{Williams:2004qba,Turyshev:2006gm,Williams:2012nc,Hofmann:2018myc}, 
and the MICROSCOPE experiment
\cite{Touboul:2017grn,MICROSCOPE:2019jix,MICROSCOPE:2022doy,Berge:2017ovy,Berge:2023sqt}.

\subsection{WEP}

The WEP experiments test the difference in accelerations between 
two test masses in an external gravity field \cite{Will:2014kxa}. 
The experimental results are usually expressed as limits 
on the E\"otv\"os parameter, which is defined as the normalized 
difference of acceleration between two bodies $A$ and $B$ 
in the same gravity field \cite{Eotvos:1922pb}
\begin{equation}
\eta_{A,B} = \left( \frac{\Delta a}{a} \right)_{A,B} 
= 2\frac{|{\bf a}_A - {\bf a}_B|}{|{\bf a}_A + {\bf a}_B|}.
\end{equation}
The WEP experiments 
also provide stringent constraints on the fifth-force.
For a Yukawa force with a Compton wavelength $\lambda=1/m_V$, 
where $m_V$ is the mediator mass, 
the E\"otv\"os parameter
in the external field of the source $S$ is 
(assuming the fifth-force is extremely small compared to gravity)
\begin{equation}
\label{eq:etaNP}
\eta_{\text{A,B}}\simeq\,
\frac{1}{4\pi G}
\left[\frac{Q_A}{m_A}-\frac{Q_B}{m_B} \right]
\frac{Q_{S}}{{m_S}} 
\, \left(1+\frac{r}{\lambda}\right)
e^{-\frac{r}{\lambda}},
\end{equation}
where 
$r$ is the distance between the experiment 
and the source $S$, 
and $Q_i$ ($m_i$) denotes the charge (mass) of the object $i$ 
with $i$ being $A$, $B$, and $S$, respectively. 
In our notation, 
the charge of a point mass in the isospin-voilating case is given by 
Eq.~\eqref{eq:Qpt}, namely $Q_{\rm pt}=f_p Z + f_n N$, 
where 
$f_p$ ($f_n$) is the mediator's coupling to proton (neutron), and 
$Z$ ($N$) is the proton (neutron) number of the point mass.
We note that Refs.~\cite{Berge:2017ovy,Wagner:2012ui} 
used a different notation: 
\begin{equation}
\eta_{\text{A,B}}\simeq\,
\alpha 
\left[ 
\left(\frac{q}{\mu}\right)_A 
- \left(\frac{q}{\mu}\right)_B 
\right]
\left( \frac{q}{\mu} \right)_S
\, \left(1+\frac{r}{\lambda}\right)
e^{-\frac{r}{\lambda}}, 
\end{equation}
where 
$\alpha$ is the parameter characterizing the strength, 
$q$ is the quantum number, 
and $\mu$ is the atomic mass in atomic units 
such that $\mu=12$ for carbon-12.  
For example, in the $U(1)_B$ case, 
$q$ is the baryon number, and   
$\alpha =  g^2/(4\pi G u^2)$, where 
$g$ is the gauge coupling, and 
$u\simeq1.66\times10^{-24}$ g is the atomic mass units.

\subsubsection{MICROSCOPE}

The MICROSCOPE experiment, 
which orbits at the altitude of 710 km, 
uses the Titanium and Platinum alloys 
as the two test masses and the Earth 
as the source of the external field 
\cite{Touboul:2017grn}.  
The recent MICROSCOPE constraints are  
$\eta_\text{Ti,Pt}=-1.5^{+3.8}_{-3.8}\times 10^{-15}$
\cite{MICROSCOPE:2022doy},
where Ti and Pt denote the two test masses. 

To compute the charges of the two test masses under the isospin mediator, 
we first note that the sizes of the test masses are small 
compared to the mediator wavelength of interest, 
and we thus neglect the effects due to their sizes. 
Table (1) of \cite{Berge:2017ovy} provides 
the $B/\mu$ and $(B-L)/\mu$ values for Ti and Pt. 
Because $B-L=Z$ and $B=Z+N$, we find that for the Ti and Pt alloys:  
$(Z/\mu)_\text{Ti}=0.40358$, 
$(Z/\mu)_\text{Pt}=0.46061$, 
$(N/\mu)_\text{Ti}=0.59668$, and 
$(N/\mu)_\text{Pt}=0.54044$. 
The charge-to-mass ratio for the Ti and Pt alloys 
can then be computed. 
For example, the charge-to-mass ratio for Ti is 
$(Q/m)_{\text{Ti}} = u^{-1}(Q/\mu)_{\text{Ti}}$ where 
$(Q/\mu)_{\text{Ti}}=f_p(Z/\mu)_\text{Ti}
+f_n(N/\mu)_\text{Ti}$.

We next compute the charge of the source (the Earth),   
denoted as $Q_E$. 
Following Ref.~\cite{Berge:2017ovy}, 
we consider a simple Earth model, 
where the Earth consists of a core and a mantle, each with a constant density. 
The charge-to-mass ratio of the Earth is 
\begin{equation}
\label{eq:earth:charge}
    \frac{Q_E}{m_E}
    =\frac{Q^{\rm pt}_c}{m_E}
    \Phi\left(\frac{R_c}{\lambda}\right)
    + \frac{Q^{\rm pt}_m}{m_E}
    \frac{\lambda^{3}}{R_E^{3}-R_c^{3}}
    \left[
    \beta\left(\frac{R_E}{\lambda}\right)-
    \beta\left(\frac{R_c}{\lambda}\right)
    \right], 
\end{equation}
where 
$m_E$ is the Earth mass, 
$\beta(x) \equiv x^3 \Phi(x)$,
$R_c$ ($R_E$) is the radius of the core (Earth), 
$Q^{\rm pt}_c$ ($Q^{\rm pt}_m$) 
is the charge of the core (mantle).\footnote{We note that 
the expression of $Q_E$ given in equation 7 of 
Ref.~\cite{Berge:2017ovy} is incorrect. We find that 
for $\lambda \lesssim 10^6$ m, 
the limits compute via Eq.~\eqref{eq:earth:charge} are 
$\sim 25\%$ higher than equation 7 of Ref.~\cite{Berge:2017ovy}.}
In our analysis, we use 
$R_c=3480$ km and $R_E=6371$ km \cite{Dziewonski:1981xy}.
We compute $Q^{\rm pt}_c$ and $Q^{\rm pt}_m$ via 
\begin{align}
\label{eq:core:charge}
   \frac{Q^{\rm pt}_c}{m_E} 
   & = 
   \frac{m_c}{m_E}\frac{1}{u}
\left(\frac{Q^{\rm pt}}{\mu}\right)_c 
=\frac{m_c}{m_E}\frac{1}{u}
\left[
f_p \left(\frac{Z}{\mu}\right)_{\rm Fe}  
+ f_n \left(\frac{N}{\mu}\right)_{\rm Fe}  
\right], 
\\ 
\label{eq:mantle:charge}
   \frac{Q^{\rm pt}_m}{m_E}
    & = 
   \left( 1 - \frac{m_c}{m_E} \right) 
   \frac{1}{u}
\left(\frac{Q^{\rm pt}}{\mu}\right)_m 
=
\left( 1 - \frac{m_c}{m_E} \right) 
\frac{1}{u}
\left[
f_p \left(\frac{Z}{\mu}\right)_{\text{SiO}_2}  
+ f_n \left(\frac{N}{\mu}\right)_{\text{SiO}_2}  
\right], 
\end{align}
where $m_c$ is the mass of the core, 
and we have assumed that 
the Earth core (mantle) consists of only Fe (SiO$_2$).
By averaging the isotopes on Earth as
given in Table \ref{tab:isotopeE}, 
we obtain that 
$\left(Z/\mu\right)_\text{Fe} \simeq 0.4656$,  
$\left(N/\mu\right)_\text{Fe} \simeq 0.5356$, 
$\left(Z/\mu\right)_{\text{SiO}_2} \simeq 0.4993$, and 
$\left(N/\mu\right)_{\text{SiO}_2} \simeq 0.5013$.
We use the Preliminary reference Earth model (PREM) \cite{Dziewonski:1981xy} 
to obtain $m_c/m_E \simeq 32.32\%$. 

The proton (neutron) number of the Earth can 
be obtained by setting 
$f_p=1$ \& $f_n=0$ ($f_n=1$ \& $f_p=0$) in 
Eqs.~(\ref{eq:earth:charge}, \ref{eq:core:charge}, \ref{eq:mantle:charge}). 
The proton and neutron numbers of the Earth 
as a function of the mediator mass are shown 
Fig.~\ref{fig:Qmv}: 
For the mediator mass $\gtrsim 10^{-14}$ eV 
(corresponding to $\lambda \lesssim 2 \times 10^4$ km), 
the proton and neutron numbers of the 
Earth start to deviate from the long-wavelength limit.

\begin{table}[htbp]
\centering    
\renewcommand{\arraystretch}{1.1}
\begin{tabular}{|c|c|c|c|}
\hline 
Element & Isotope & Abundance ($\%)$ & $\mu $ \\
\hline 
 \multirow{3}{*}{O} & $^{16}$O &  99.7621 & 15.995 \\
\cline{2-4} 
   & $^{17}$O &  0.0379 & 16.999 \\
\cline{2-4} 
   & $^{18}$O &  0.2000 & 17.999 \\
\hline 
 \multirow{3}{*}{Mg} & $^{24}$Mg & 78.99 & 23.985 \\
\cline{2-4} 
   & $^{25}$Mg & 10.00 & 24.986 \\
\cline{2-4} 
   & $^{26}$Mg & 11.01 & 25.983 \\
\hline 
 Al & $^{27}$Al & 100.0 & 26.982 \\
\hline 
 \multirow{3}{*}{Si} & $^{28}$Si & 92.2297 & 27.977 \\
\cline{2-4} 
   & $^{29}$Si & 4.6832 & 28.976 \\
\cline{2-4} 
   & $^{30}$Si & 3.0872 & 29.974 \\
\hline 
 \multirow{6}{*}{Ca} & $^{40}$Ca & 96.941 & 39.963 \\
\cline{2-4} 
   & $^{42}$Ca & 0.647 & 41.959 \\
\cline{2-4} 
   & $^{43}$Ca & 0.135 & 42.959 \\
\cline{2-4} 
   & $^{44}$Ca & 2.086  & 43.955 \\
\cline{2-4} 
   & $^{46}$Ca &  0.004 & 45.954 \\
\cline{2-4} 
   & $^{48}$Ca & 0.187 & 47.953 \\
\hline 
 \multirow{4}{*}{Fe} & $^{54}$Fe &  5.845 & 53.940 \\
\cline{2-4} 
   & $^{56}$Fe & 91.754 & 55.935 \\
\cline{2-4} 
   & $^{57}$Fe & 2.119 & 56.935 \\
\cline{2-4} 
   & $^{58}$Fe & 0.282 & 57.933 \\
 \hline
\end{tabular}
\caption{The isotopes of various elements, 
the natural abundance (mole-fraction) of the isotopes in the Earth
\cite{10.1063/1.556031}, 
and the atomic mass \cite{elements}.}
\label{tab:isotopeE}
\end{table}

\begin{table}[htbp]
\centering    
\renewcommand{\arraystretch}{1.1}
\begin{tabular}{|c|c|c|c|}
\hline 
Element & Isotope & Abundance ($\%)$ & $\mu $ \\
\hline 
\multirow{2}{*}{H} & $^{1}$H & 99.998 & 1.008 \\
\cline{2-4} 
& $^{2}$H & 0.002 & 2.014 \\
\hline 
\multirow{2}{*}{He} & $^{3}$He &  0.0166 & 3.016 \\
\cline{2-4} 
& $^{4}$He & 99.9834 & 4.003 \\
\hline
\end{tabular}
\caption{The isotopes of hydrogen and helium, 
the natural abundance (mole-fraction) of the isotopes
in the solar system \cite{Asplund:2009fu}, 
and the atomic mass \cite{elements}.}
\label{tab:isotopeS}
\end{table}

\subsubsection{LLR experiments}
\label{sec:LLR}

The LLR experiments test both weak and strong equivalence principles, 
with the measurements of round trip
travel times of short laser pulses between 
observatories on the Earth and retro-reflectors
on the Moon \cite{Hofmann:2018myc}. 
LLR provides constraints on 
the Earth-Moon differential
acceleration in the field of the Sun: 
\cite{Hofmann:2018myc}:
$\eta_\text{Earth,Moon}=-3^{+5}_{-5}\times 10^{-14}$. 
To analyze the constraints on the isospin-violating mediator, 
we need to compute the charges of the Earth, the Moon, 
and the Sun. 
Because the Earth and the Moon orbit the Sun
at a distance of $\simeq 1.5\times 10^8$ km (1 AU), 
which is very large compared to the radius of the Sun
$R_\odot\simeq 7\times 10^5$ km (as well as 
the radius of the Earth  
$R_\oplus$ and 
the radius of the Moon
$R_\text{Moon}$), 
the effects due to their sizes are negligible. 
The calculation of the charge of the Earth 
is the same as the MICROSCOPE experiment.

We next compute the charge of the Sun. 
The Sun is mainly composed of hydrogen and helium, 
with mass fractions of $73.81\%$ and $24.85\%$ \cite{Asplund:2009fu}, 
and the abundance of the various isotopes shown in Table \ref{tab:isotopeS}.
For the heavy elements, which are small in the mass fraction of the Sun, 
we take $(Z/\mu)=(N/\mu)\simeq 0.5$. 
We thus obtain $(Z/\mu)_\odot \simeq 0.8632$
and $(N/\mu)_\odot \simeq 0.1309$ for the Sun.

We next compute the charge of the Moon. 
We adopt the Lunar Primitive Upper
Mantle (LPUM) model \cite{longhi2006petrogenesis} 
for the Moon, in which the major compositions include 
$46.1\%$ SiO$_2$, 
$38.3\%$ MgO,
$7.62\%$ FeO, 
$3.93\%$ Al$_2$O$_3$, and 
$3.18\%$ CaO. 
We assume that the natural abundance of the various isotopes  
are the same as the Earth, as given in given in Table \ref{tab:isotopeS}. 
By averaging the various isotopes for the five major compositions 
and taking $(Z/\mu)=(N/\mu)\simeq 0.5$ for all other minor compositions, 
we obtain $(Z/\mu)_\text{Moon}=0.4958$ 
and $(N/\mu)_\text{Moon}=0.5048$ 
for the moon.

\subsubsection{EW experiment}

The EW experiment utilizes a remarkable sensitive
torsion balance to test the WEP violation, 
which has achieved a precision of $10^{-13}$ 
\cite{Wagner:2012ui,Schlamminger:2007ht}. 
The EW experiment compares different pairs of test bodies, including 
the beryllium-aluminum pair 
and the beryllium-titanium pair; 
it also uses different attractors, including 
geocenter, the Sun, and the galactic center \cite{Wagner:2012ui}.
This enables it to probe the composition-dependent fifth force 
\cite{Wagner:2012ui,Fischbach:2020kwt}. 
The EW constraints on the baryon number
case
and the $B-L$ case are given in 
Refs.~\cite{Schlamminger:2007ht,Wagner:2012ui}.

\subsection{Asteroids and Planets}

Trajectories of asteroids and planets (hereafter AP) 
around the Sun 
offer an ideal testing ground for the laws of physics, 
where deviations from general relativity predictions 
can be used to probe new physics.
In the presence of a Yukawa type interaction,
the equation of motion of the asteroids and planets 
in the gravitational field of the Sun is given by 
\cite{KumarPoddar:2020kdz,Tsai:2021irw,Tsai:2023zza}:
\begin{equation}
    \label{eq:equationsofmotion}
    \frac{{\rm d}^2 u}{{\rm d}\varphi^2} + u = 
    \frac{GM_\odot}{L^2}+\frac{3GM_\odot}{c^2} u^2 + 
    \alpha\frac{G M_\odot}{L^2} 
    \left( 1+\frac{1}{\lambda u} \right) 
    e^{-\frac{1}{\lambda u}}\,,
\end{equation}
where $u=1/r$ with $r$ being the distance between these celestial objects, 
$\varphi$ is the azimuthal angle of the motion, 
$L$ denotes the orbital angular momentum per unit mass, 
$\lambda=1/m_V$ is the Compton wavelength of the mediator, 
and $\alpha$ is given by Eq.~\eqref{eq:alpha} 
with the two objects being the Sun and the asteroid/planet.
The first, second and third terms on the right-hand side 
of Eq.~\eqref{eq:equationsofmotion} 
describe the effects of Newtonian physics, GR corrections, and
the fifth force, respectively.

The data of asteroids \cite{Tsai:2021irw} \cite{Tsai:2023zza}
and planets \cite{Talmadge:1988qz,KumarPoddar:2020kdz}
in the solar system have been used to place constraints on 
new physics, including the fifth force. 
In our analysis 
we adopt the upper bound on $\alpha$ in figure 1 of \cite{Tsai:2023zza} which combines the results 
in Refs.~\cite{KumarPoddar:2020kdz,Tsai:2021irw,Tsai:2023zza}. 
To compute the constraints 
on the isospin-violating mediator, 
we calculate the charge of the Sun as in 
section \ref{sec:other:constraints}; 
for the asteroids and planets, which 
mostly consist of heavy elements, 
we use $(Z/\mu)_{\rm AP}=(N/\mu)_{\rm AP} \simeq 0.5$. 
Because trajectories of asteroids and planets are 
sensitive to the fifth force mediator in the wavelength $\simeq (0.1,\,10)$ AU, we do not consider the effects 
due to the finite size of the Sun, asteroids, and planets. 
For heavy mediator mass $\gtrsim 10^{-17}$ eV, 
the dominant AP constraints come from Ref.~\cite{Tsai:2021irw}, 
which, however, only provides upper bound on $\alpha$ 
for $m_V \lesssim 2\times 10^{-16}$ eV. 
To compare with the GW170917 limits, 
we extrapolate the AP constraints with 
an exponential factor of $e^{-a\,m_V}$ 
\cite{KumarPoddar:2020kdz}, 
and  fit the coefficient $a$ using the 
constraints for $m_V < 2\times 10^{-16}$ eV. 
The results is shown in 
Figs.~\ref{Fig:Constraints-U-B}-\ref{Fig:Constraints-fixed-coupling}.

\subsection{BHSR}

The energy and angular momentum of a black hole 
can be transferred into its surrounding ``cloud'' 
that consists of ultralight bosonic fields, 
through the process known as the BHSR
\cite{Brito:2015oca}. 
The BHSR relies on only the gravitational interaction, 
thereby making it applicable to a variety of ultralight bosonic fields.
Moreover, because the ultralight bosonic fields can be spontaneously produced, 
the BHSR does not require a preexisting abundance of the bosonic fields 
\cite{Baryakhtar:2017ngi}. 
The BHSR constraints 
on ultralight weakly-coupled spin-1 particles have been 
analyzed in Ref.~\cite{Baryakhtar:2017ngi}: 
the measurements of rapidly spinning BHs 
in X-ray binaries exclude vector particles in the mass range of 
$(5 \times 10^{-14},\, 2 \times 10^{-11})$ eV, 
and the spin measurements of supermassive BH 
exclude vector particles in the mass range of 
$(6 \times 10^{-20},\, 2 \times 10^{-17})$ eV
(with a lower confidence level). 
Note that the BHSR constraints are analyzed under the assumption that 
the ultralight bosonic field does not possess a significant particle 
interaction with itself or with other particles. 
The presence of such interactions can potentially invalidate 
the BHSR constraints; 
see e.g., Refs.~\cite{Siemonsen:2022ivj,Cannizzaro:2022xyw} 
and also Refs.~\cite{Baryakhtar:2020gao,Unal:2020jiy,Mehta:2020kwu} 
for BHSR constraints on self-interacting axions.

\section{Results}
\label{sec:results}

In this section, we derive constraints on the 
isospin-violating mediator based on GW170817. 
There are three NP parameters in the model: 
the mediator mass $m_V$, 
the coupling to neutrons $f_n$, and 
the coupling to protons $f_p$. 
We carry out MCMC scans for fixed $m_V$ values, resulting in 
two NP parameters for each mediator mass. 
Because the 2D parameter space defined by 
$f_p$ and $f_n$ for each $m_V$ 
is uniquely determined by the 2D space spanned by 
$\alpha$ and $\gamma$, 
we perform MCMC analysis with $\alpha$ and $\gamma$ 
as the free parameters. 
This approach offers advantages: 
Firstly, $\alpha$ and $\gamma$ hold more direct physical significance 
in the context of GW phenomenology. 
Secondly, the obtained limits on $\alpha$ and $\gamma$ can be 
readily applied to other new light mediators.

We next discuss our method to analyze 
the GW170817 constraints 
in the parameter space of the 
isospin-violating mediators. 
We discuss the bound related 
to the $\alpha$ parameter first. 
In our analysis we choose a 
fixed ratio of $f_p/f_n$ 
and use Eq.~\eqref{eq:alpha} 
to compute the $f_n$ value from the 
$\alpha$ value (by using the 
$m_1$ and $m_2$ values)
for each MCMC sampling point. 
Thus we obtain a model point 
in the new 11D parameter space 
where $\alpha$ is replaced by $f_n$. 
We then magnilize all the other
10 parameters to obtain the 
marginalized posterior on $f_n$. 
The bound related to the $\gamma$ parameter is 
done in a similar way.

\begin{figure}[htbp]
\begin{centering}
\includegraphics[width=0.49 \textwidth]{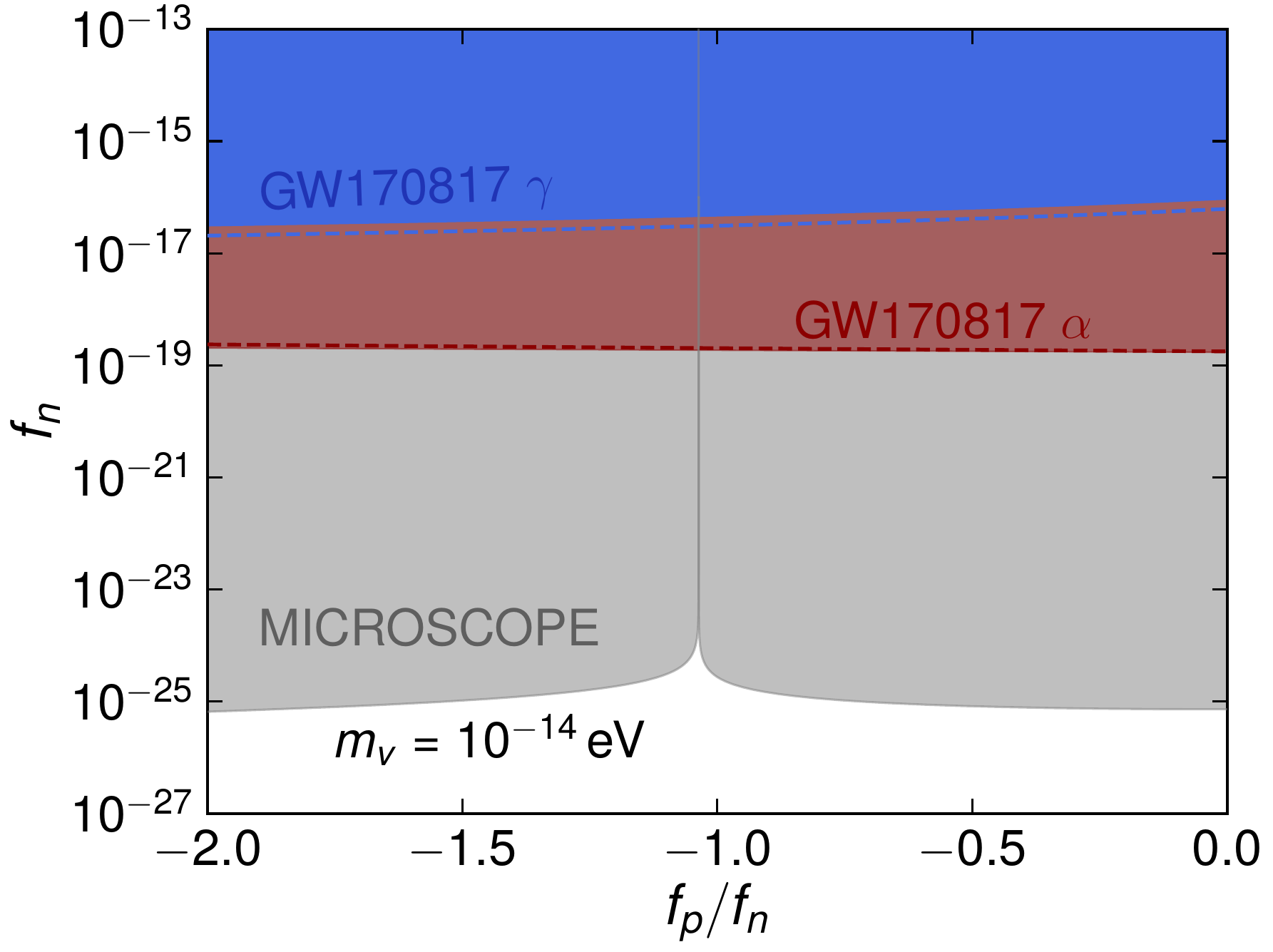}
\includegraphics[width=0.49 \textwidth]{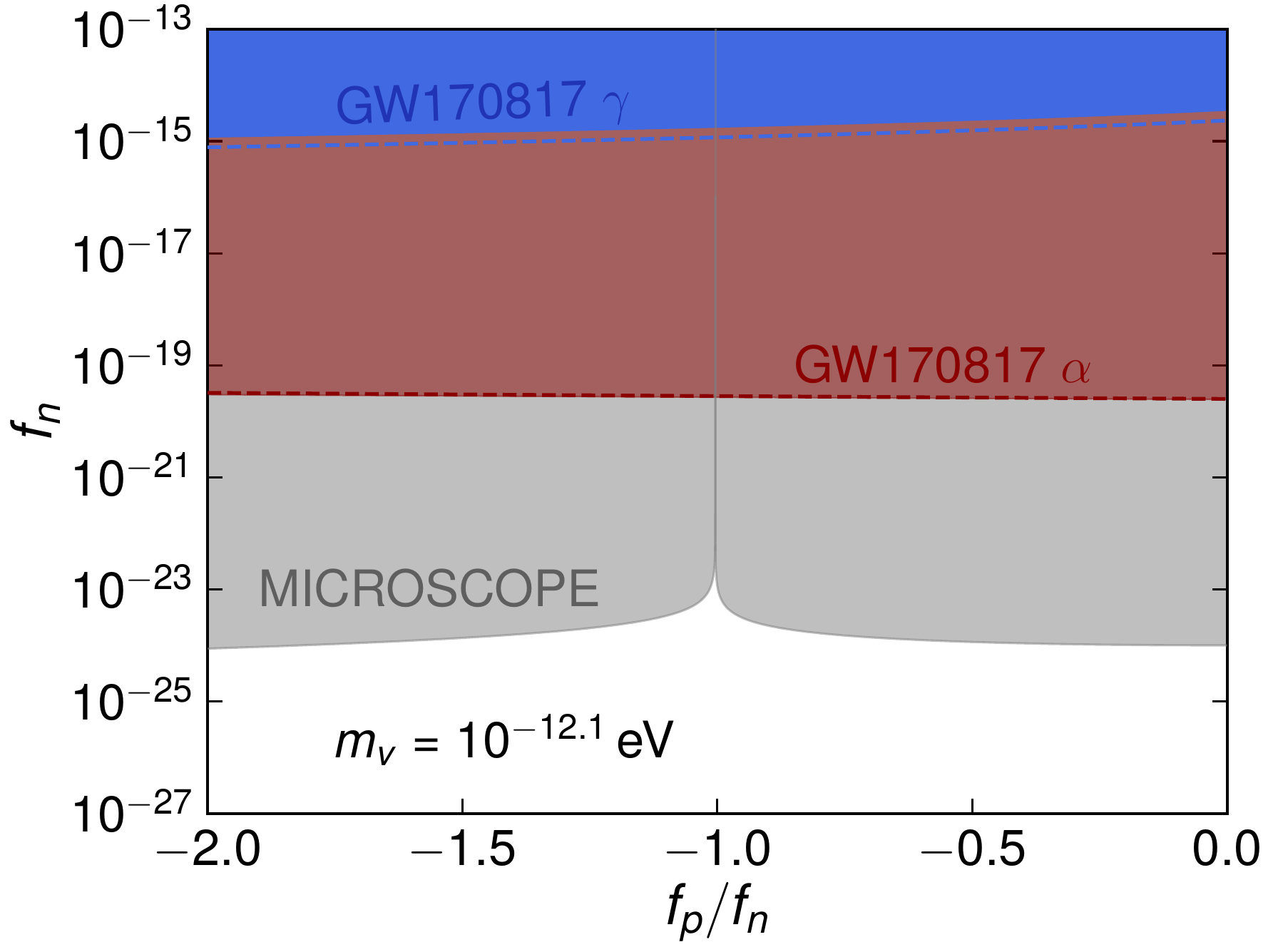}
\caption{GW170817 constraints
($3\sigma$) on the parameter space spanned by 
$f_p/f_n$ and $f_n$, for 
$m_V=10^{-14}$ eV ({\bf left panel}) and
$m_V=10^{-12.1}$ eV ({\bf right panel}). 
Constraints from the $\alpha$ and $\gamma$ 
bounds are shown as the red and blue shaded
regions, respectively, 
with the default EoS, BSk24 \cite{Pearson:2018tkr}; 
in comparison, we also show the upper bounds  
with the BSk22 EoS (dashed)
\cite{Pearson:2018tkr}.
Also shown are 
constraints from the MICROSCOPE
experiment (gray) \cite{Berge:2023sqt}.
The $m_V=10^{-12.1}$ eV case is also constrained by 
the black hole superradiance constraints \cite{Baryakhtar:2017ngi}.
}
\label{Fig:couplings-ratio}
\end{centering}
\end{figure}

Fig.~\ref{Fig:couplings-ratio} shows the GW170817 constraints 
on the parameter space spanned by $f_n$ and $f_p/f_n$ 
for two mediator masses: 
$m_V = 10^{-14}$ eV (left panel figure), and 
$m_V = 10^{-12.1}$ eV (right panel figure); 
for both cases, the constraints due to $\alpha$ 
are stronger than $\gamma$. 
We also compare the GW170817 constraints 
with the MICROSCOPE constraints: 
for both panel figures in Fig.~\ref{Fig:couplings-ratio}, 
the MICROSCOPE constraints 
\cite{MICROSCOPE:2022doy,Berge:2023sqt} 
are several orders of magnitude stronger 
than the GW170817 constraints, 
except in the vicinity of $f_p/f_n\simeq -1$, 
where the charge of the Earth becomes zero. 
We note that
the $m_V = 10^{-12.1}$ eV case is 
also constrained by black hole superradiance 
\cite{Baryakhtar:2017ngi}.
In addition to the default EoS, BSk24, 
we also compute the upper bound on $f_n$ in Fig.~\ref{Fig:couplings-ratio} 
by using the $\alpha$ and $\gamma$ constraints, 
with an alternative EoS, BSk22 \cite{Pearson:2018tkr}. 
We find that the differences between BSk24 and BSk22 are small.

\begin{figure}[htbp]
\begin{centering}
\includegraphics[width=0.45 \textwidth]{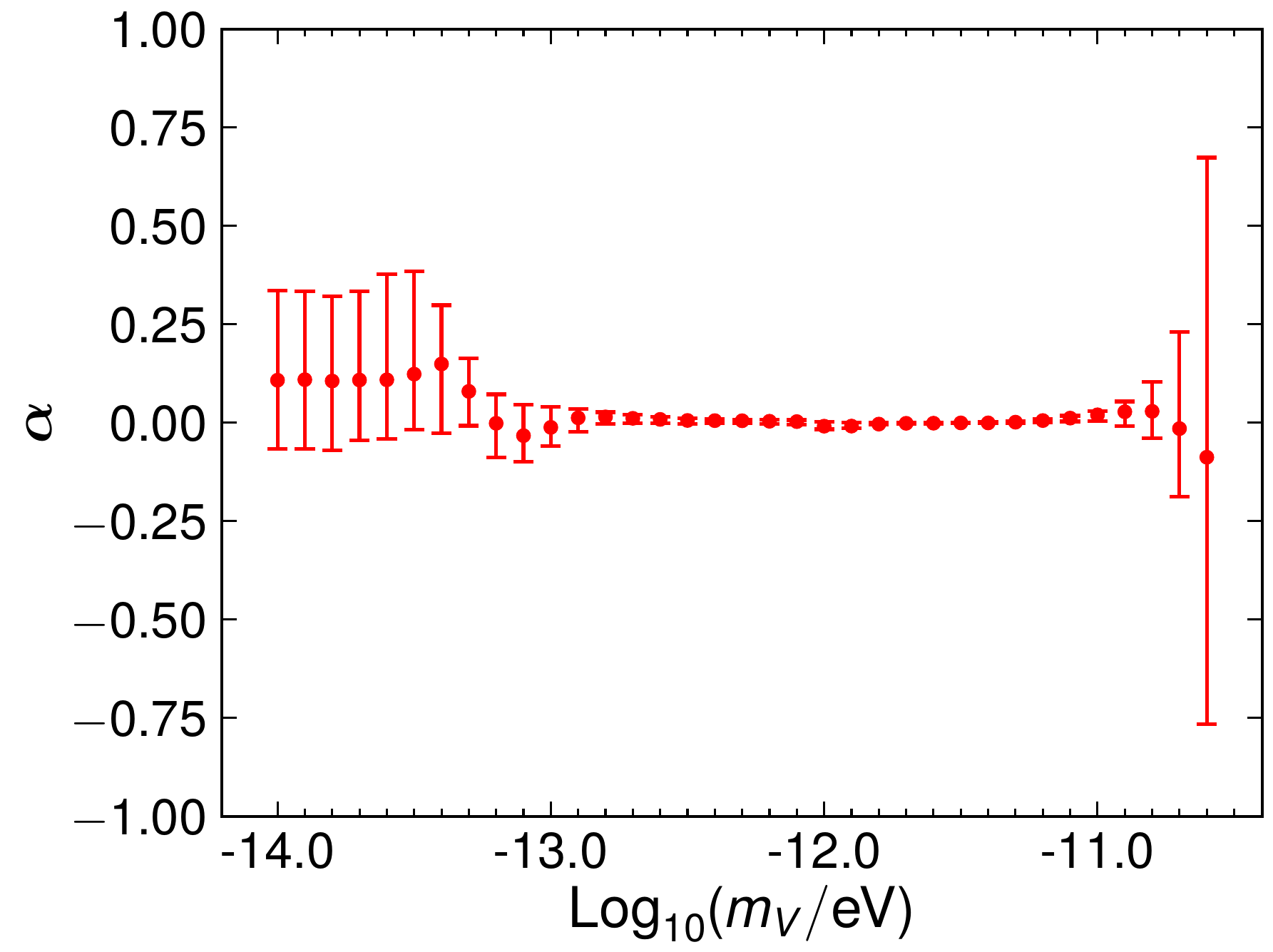}
\includegraphics[width=0.45 \textwidth]{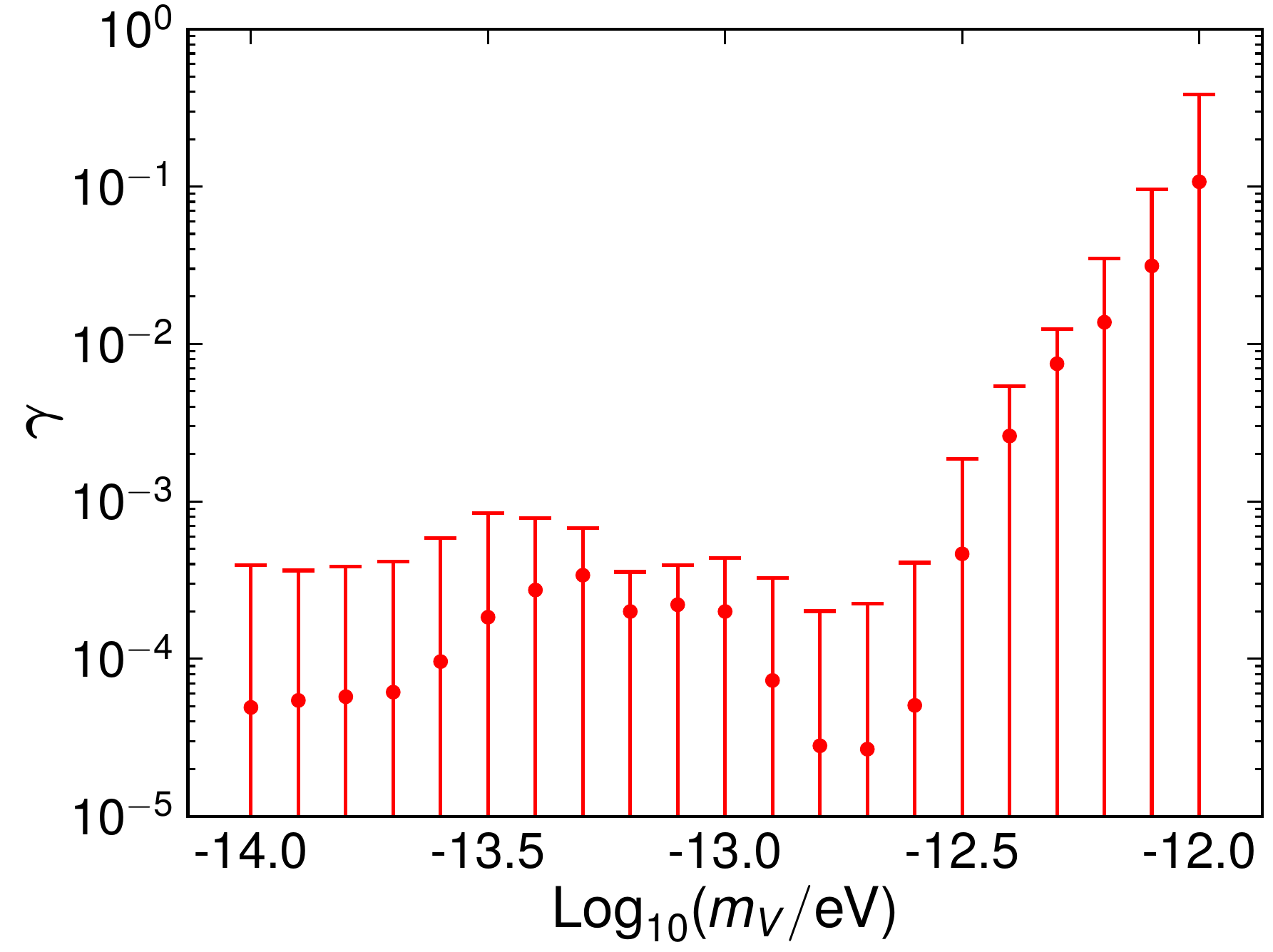}
\caption{The $3 \sigma$ bounds on $\alpha$ 
({\bf left panel})  
and $\gamma$ ({\bf right panel}) for different mediator masses 
from the BNS merger event
GW170817, 
where the dots indicate the medians. 
We compute the limits, by using the GW170817 data from Ref.~\cite{LIGOScientific:2019lzm}.
}
\label{fig:results}
\end{centering}
\end{figure}

Moreover, we provide in Fig.~\ref{fig:results}
the bounds on the $\alpha$ and $\gamma$ parameters
to show the detectability of the fifth force
in the GW170817 event.
To obtain bounds on $\alpha$, 
we compute the marginalized posterior on $\alpha$, 
by marginalizing over all other 10 parameters in the 
posteriors from the MCMC sampling. 
The left panel figure of Fig.~\ref{fig:results} shows the 
$3\,\sigma$ 
($99.7\%$) 
credible regions on $\alpha$. 
Unlike $\alpha$, which can take either positive or negative 
value, $\gamma$ is always positive. 
Thus, the 3 $\sigma$ bound on $\gamma$ only has an upper bound, 
which is defined such that the total probability 
in the parameter range from zero to the upper bound is $99.7\%$. 
The right panel figure of Fig.~\ref{fig:results} shows the 
$3\,\sigma$ 
($99.7\%$) 
upper bounds on $\gamma$.
We do not show in Fig.~\ref{fig:results} the 
constraints on $\alpha$ ($\gamma$) 
for $m_V \gtrsim 10^{-10.5}$ ($10^{-12}$) eV, 
where the constraints on $\alpha$ ($\gamma$) become of order one 
so that perturbative calculations can no longer be trusted.

We note that 
the bound on $\alpha$ can be used 
to place bounds in some other ultralight mediator models. 
We find that the bound on $\alpha$ in Fig.~(\ref{fig:results}) 
leads to a limit on $f_n$ (with
$m_1=1.46\,M_\odot$ and $m_2=1.27\,M_\odot$)
that is only $\lesssim 2\%$ different 
from that in Fig.~\ref{Fig:couplings-ratio}. 
However, 
the bound on $\gamma$ should be used in caution. 
We find that the bound on $\gamma$ in Fig.~\ref{fig:results} 
leads to a limit on $f_n$ that is about two orders of magnitudes stronger 
than in Fig.~\ref{Fig:couplings-ratio}. 
This is largely due to the fact that $\gamma$, 
computed via Eq.~\eqref{eq:gamma}, 
is proportional 
to the difference of the charge-to-mass ratios of the two NSs in the 
GW170817 event, which, however, have similar masses. 
Thus, the interpretation of $\gamma$ is very sensitive to 
the masses of the NSs.

We next consider two special cases 
in the parameter space of the iso-spin violating 
model: 
(1) 
the baryon number case in which $f_p=f_n$; 
(2) the case where the Earth charge is zero, 
namely $Q_E=0$.

\subsection{The baryon number case}

\begin{figure}[htbp]
\begin{centering}
\includegraphics[width=0.65 \textwidth]{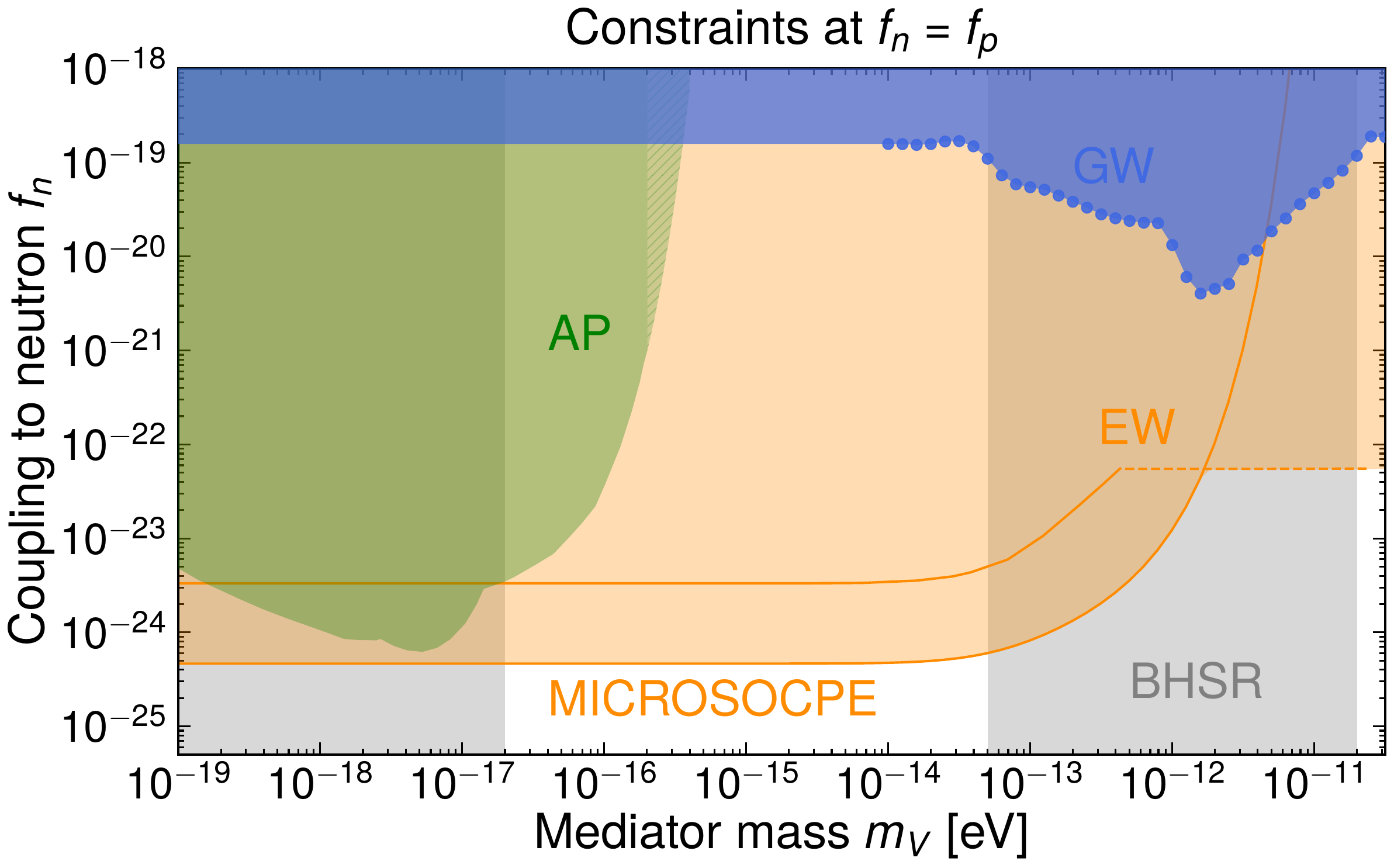}
\caption{The $3 \sigma$ constraints on
the coupling $f_n$ from the BNS merger event 
GW170817 {(denoted as GW)} (blue),
for the $f_n=f_p$ case. 
The blue dots indicate the actual model points 
that are given in Fig.~\ref{fig:results}. 
We extrapolate the GW constraint
from $m_V=10^{-14}$ eV to the massless limit,
as the GW170817 data cannot distinguish
the fifth force signals for such a low mass
mediator.
Other constraints are also shown here
as shaded regions: 
asteroids and planets (denoted as AP) data (green)
\cite{KumarPoddar:2020kdz,Tsai:2021irw, Tsai:2023zza}, 
MICROSCOPE (orange) 
\cite{Berge:2023sqt,MICROSCOPE:2022doy}, 
EW (orange) 
\cite{Schlamminger:2007ht}, 
black hole superradiance (denoted as BHSR) (gray)
\cite{Baryakhtar:2017ngi}. 
Note that the AP limits for  
$m_V\gtrsim2\times10^{-16}$ eV are extrapolated 
from Ref.~\cite{Tsai:2021irw}.
}
\label{Fig:Constraints-U-B}
\end{centering}
\end{figure}

Here we consider the baryon number case, namely $f_p=f_n$. 
For each mediator mass $m_V$, 
we compute the constraints on
$f_n$ with the same method as in Fig.~\ref{Fig:couplings-ratio}.
The obtained GW170817 constraints are shown in 
Fig.~\ref{Fig:Constraints-U-B}. 

In our analysis we have carefully taken into account 
the effects due to the finite sizes of the NSs. 
We find that for heavy mediator mass 
in the range of $m_V \simeq (2-3)\times10^{-11}$ eV, 
the NS charge is a factor of $\simeq(13-34)$\% larger 
than the naive calculation in which the NS is treated 
as a point charge. 
This then leads 
to an upper bound on the coupling that is smaller
by the same factor. 
We note that because the effects due to the finite sizes of the NSs 
increase significantly with the mediator mass,
as shown in the left panel figure of Fig.~\ref{fig:Qmv}, 
it alleviates somewhat the exponential suppression of 
the Yukawa force on heavy mediators. 
We also note that in our analysis the parameter 
region where the effects due to the finite sizes of the NSs 
become significant coincides with the parameter 
region where the perturbation calculations start to fail. 
However, in certain cases, 
the effects due to the finite sizes of the NSs 
may become significant in regions 
where perturbation calculations remain reliable. 
For example, a future BNS merger that occurs with a distance 
much smaller than the GW170817 event may provide a much stronger 
constraint on $\alpha$ 
such that the constraints at
$\gtrsim 3\times 10^{-11}$ eV can still
be obtained via perturbation calculations.

Fig.~\ref{Fig:Constraints-U-B} also shows
the other dominant experimental constraints
in the parameter space of interest,
including those from
the EW experiment \cite{Schlamminger:2007ht},
the MICROSCOPE experiment
\cite{MICROSCOPE:2022doy,Berge:2023sqt},
the asteroids and planets (AP)
\cite{KumarPoddar:2020kdz,Tsai:2021irw,Tsai:2023zza},
and the BHSR \cite{Baryakhtar:2017ngi}. 
We find that in the baryon number case, 
the GW170817 constraints are weaker than 
the other experimental constraints 
for the mediator mass $m_V \lesssim 10^{-11}$ eV.

\subsection{The $Q_E=0$ case}
\label{sec:qezero}

Here we consider the case where 
the charge of Earth is zero, namely $Q_E=0$.
This particular case is of great interest, as 
it is not constrained by the MICROSCOPE and EW 
experiments, which use Earth as the source of
the external field.

\begin{figure}[htbp]
\begin{centering}
\includegraphics[width=0.45 \textwidth]{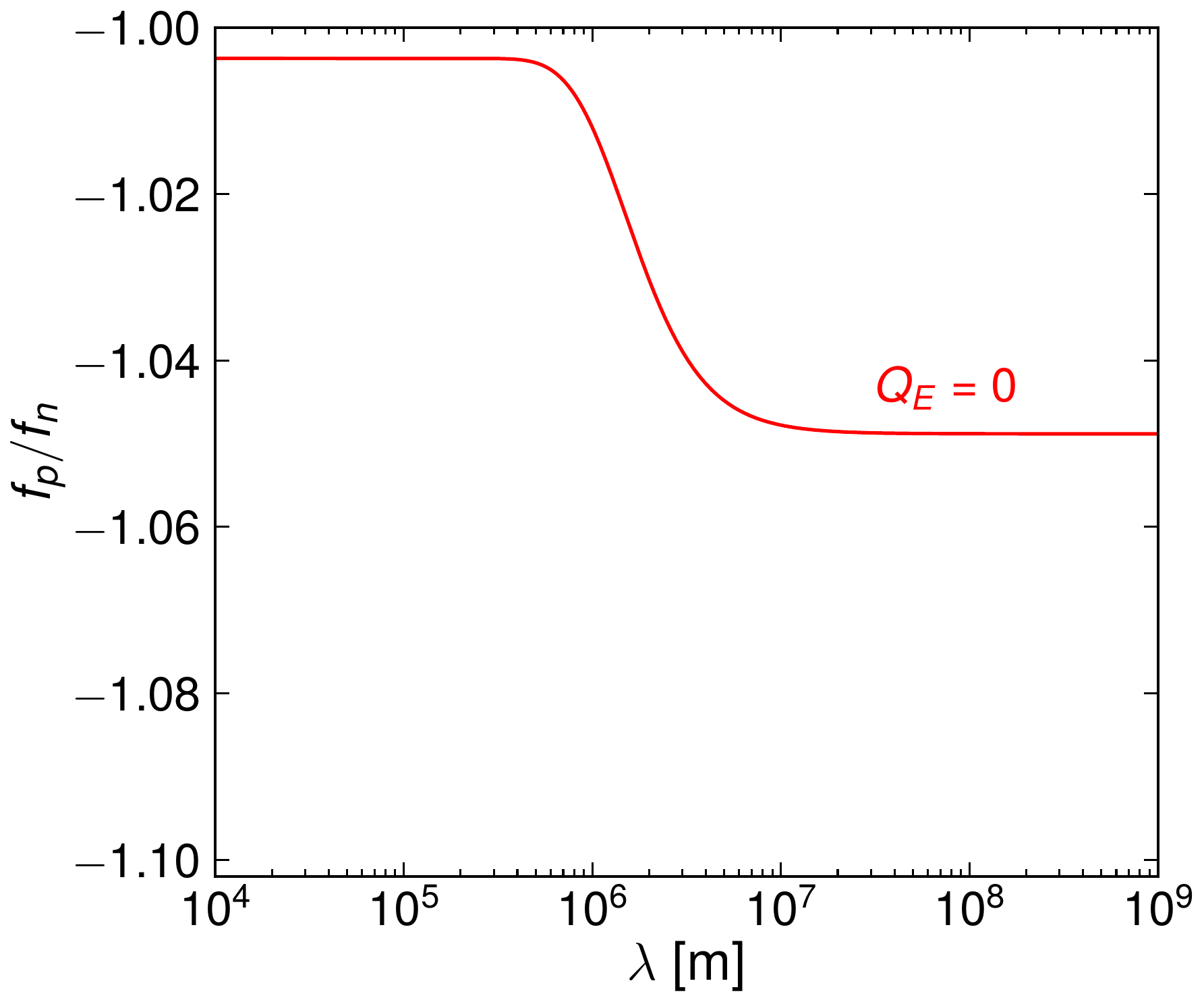}
\caption{The $Q_E=0$ curve in the parameter space 
spanned by $f_p/f_n$ and $\lambda=1/m_V$.}
\label{Fig:zero-Q}
\end{centering}
\end{figure}

To compute the GW170817 constraints on the $Q_E=0$ case, 
we first determine for each mediator mass 
the ratio $f_p/f_n$ that satisfies $Q_E=0$ 
by using Eq.~\eqref{eq:earth:charge}:  
the $Q_E=0$ curve in the parameter space 
spanned by $f_p/f_n$ and $\lambda=1/m_V$ is shown 
in Fig.~\ref{Fig:zero-Q}.
We note that the $f_p/f_n$ value that satisfies $Q_E=0$ 
is $m_V$-dependent, due to the effects of the finite size of the Earth.
Because of the $Q_E=0$ condition, the 3D parameter space 
($m_V$, $f_p$, $f_n$) is now reduced to a 2D parameter space, 
for which we use ($m_V$, $\alpha$). 
Thus, we perform MCMC runs with $m_V$ and $\alpha$ as 
free parameter, but with $\gamma$ given by 
\footnote{We note that 
MCMC runs in a constrained 2D parameter space 
instead of the 3D parameter space, ($m_V$, $f_p$, $f_n$), 
should also be carried out for the $f_p=f_n$ case to accurately 
determine the GW170817 constraints. 
In our analysis we have neglected such MCMC runs 
because they only produce shifts on $f_n$ by $\lesssim 50\%$,  
and because the GW170817 constraints in the $f_p=f_n$ case 
are much weaker than other experimental constraints.}
\begin{equation}
    \gamma=\alpha
    \left( \frac{Q_1}{m_1}\frac{Q_2}{m_2} \right)^{-1} 
    \left( \frac{Q_1}{m_1}-\frac{Q_2}{m_2} \right)^2. 
\end{equation}

\begin{figure}[htbp]
\begin{centering}
\includegraphics[width=0.65 \textwidth]{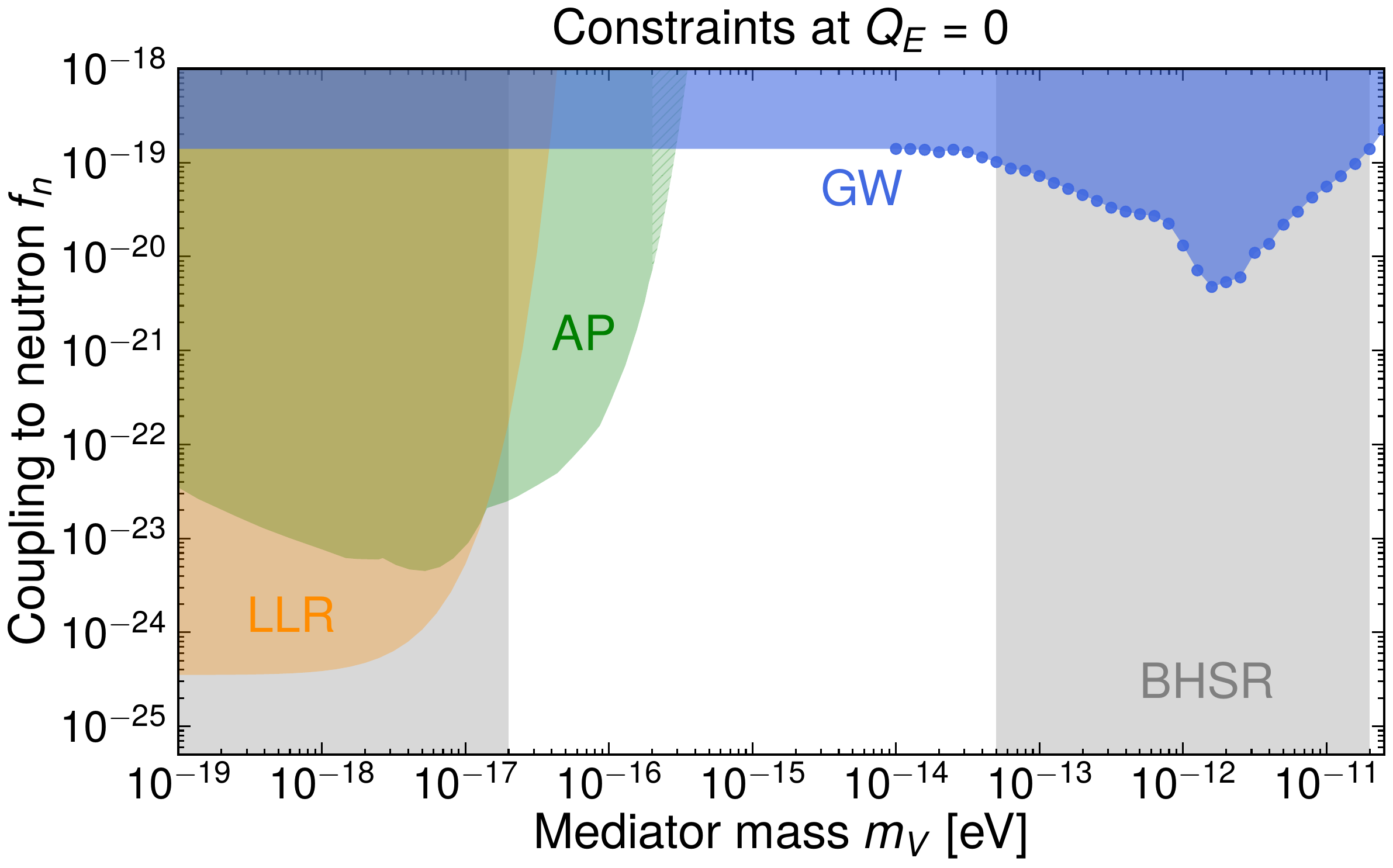}
\caption{
The $3\sigma$ constraints on
the coupling $f_n$ from the BNS merger event
GW170817 {(denoted as GW)} (blue),
for the $Q_E = 0$ case. 
The blue dots indicate the actual model points 
that are given in Fig.~\ref{fig:results}. 
We extrapolate the GW constraint
from $m_V=10^{-14}$ eV to the massless limit,
as the GW170817 data cannot distinguish
the fifth force signals for such a low mass
mediator.
Other constraints are also shown here
as shaded regions: 
asteroids and planets (denoted as AP) data (green)
\cite{KumarPoddar:2020kdz,Tsai:2021irw, Tsai:2023zza}, 
lunar laser ranging (denoted as LLR) results (orange)
\cite{Hofmann:2018myc}, 
and the black hole superradiance
(denoted as BHSR) (gray)
\cite{Baryakhtar:2017ngi}. 
Note that the AP limits for  
$m_V\gtrsim2\times10^{-16}$ eV are extrapolated 
from Ref.~\cite{Tsai:2021irw}.
}
\label{Fig:Constraints-fixed-coupling}
\end{centering}
\end{figure}

Fig.~\ref{Fig:Constraints-fixed-coupling} 
shows the GW170817 constraints on the $Q_E=0$ case. 
Because the Earth charge is zero, 
the leading constraints in Fig.~\ref{Fig:Constraints-U-B} 
from
the EW experiment \cite{Schlamminger:2007ht} 
and from 
the MICROSCOPE experiment
\cite{MICROSCOPE:2022doy,Berge:2023sqt}, 
both of which use Earth as the source of the external fields, 
are no longer present. 
The other dominant experimental constraints now 
include LLR  
\cite{Hofmann:2018myc}, 
the asteroids and planets (AP) \cite{KumarPoddar:2020kdz,Tsai:2021irw,Tsai:2023zza}, 
and the BHSR constraints \cite{Baryakhtar:2017ngi}.
We find that in the $Q_E=0$ case,
the GW170817 data provide the most stringent 
constraints in the mediator mass range 
$\simeq(3\times10^{-16},\,5\times10^{-14})$ eV.

\section{Summary}
\label{sec:suma}

We analyze the constraints on ultralight 
isospin-violating mediators from the GW170817 event, 
the first and most robust BNS merger event detected by 
the LIGO/Virgo collaboration. 
Although fifth-force experiments, such as 
MICROSCOPE and EW, usually impose more stringent 
constraints on ultralight isospin-violating mediators 
(see e.g., the constraints on the $U(1)_B$ mediator, 
a special case of the isospin-violating mediator, 
in Fig.~\ref{Fig:Constraints-U-B}), 
we find that the GW170817 event can offer leading 
constraints on the parameter space where the 
isospin-violating force is screened by the Earth 
(namely the charge of the Earth is zero). 
In such cases, the GW170817 event excludes 
the existence of ultralight isospin-violating mediators 
down to $f_n \simeq \mathcal{O}(10^{-19})$ 
and provides the most constraining limit 
in the mediator mass range of 
$\simeq(3\times10^{-16},\,5\times10^{-14})$ eV
(corresponding to the Compton wavelength range of 
$\lambda\simeq(4\times10^3\,,6\times10^5)$ km), 
as shown in Fig.~\ref{Fig:Constraints-fixed-coupling}.

To analyze the 
fifth-force-induced
GW signal, 
we compute the gravitational waveform by taking into 
account the new physics effects, including the long-range 
Yukawa force and the dipole radiation. 
We note that these two new physics effects have different
dependences on the charge-to-mass ratios of the NSs: 
the Yukawa force is proportional to
the product of the charge-to-mass ratios, while the 
dipole radiation is proportional to 
the quadratic power of their difference.
Moreover, we note that the GW170817
constraints given
by the {dipole} radiation of the light mediator are
suppressed due to the small charge difference
between the NSs; 
we anticipate that future detection of NS-BH
mergers can potentially
strengthen these constraints by orders of magnitude.

To correctly interpret the experimental data, 
we have computed the charges of the NSs by taking 
into account the effects due to their finite sizes, 
which are particularly important when 
the wavelength of the mediator becomes small compared 
to the NS radius. 
In such cases, the charge is much larger than in the 
long-wavelength limit where the NS is treated as 
a point charge. 
This enlargement of the NS charge at short wavelength 
(large mediator mass) offsets somewhat the exponential decline of 
the constraints on the mass, making the large mediator mass region 
more promising than naively expected. 
For the interpretation of other constraints, 
we have provided a simple analytic expression 
of the Earth charge, correcting a previous 
erroneous formula in Ref.~\cite{Berge:2017ovy}.

In our analysis we have used perturbation
calculations to compute the new physics effects
and restricted our attention 
to the mediator mass range of $\lesssim 10^{-11}$ eV, 
beyond which the 
perturbation calculations begin to fail. 
We note that the mass range can be extended to 
larger mediator masses, if one uses other calculations 
instead of the perturbative ones. We leave this to 
our future studies.

\begin{acknowledgments}
We thank Joel Berg\'{e}, Huai-Ke Guo, and
Yong-Heng Xu 
for discussions and correspondence. 
We thank Alexander Nitz and Collin D.\ Capano 
for helpful correspondence on PyCBC.
The work is supported in part by the 
National Natural Science Foundation of China under Grant Nos.\ 12275128 and 12147103.

This research has made use of data or software obtained from 
the Gravitational Wave Open Science Center (gwosc.org), a 
service of LIGO Laboratory, the LIGO Scientific Collaboration, 
the Virgo Collaboration, and KAGRA. LIGO Laboratory and 
Advanced LIGO are funded by the United States National Science
Foundation (NSF) as well as the Science and Technology 
Facilities Council (STFC) of the United Kingdom, the Max-Planck-
Society (MPS), and the State of Niedersachsen/Germany for 
support of the construction of Advanced LIGO and construction 
and operation of the GEO600 detector. Additional support for 
Advanced LIGO was provided by the Australian Research Council. 
Virgo is funded, through the European Gravitational Observatory 
(EGO), by the French Centre National de Recherche Scientifique 
(CNRS), the Italian Istituto Nazionale di Fisica Nucleare 
(INFN) and the Dutch Nikhef, with contributions by institutions 
from Belgium, Germany, Greece, Hungary, Ireland, Japan, Monaco, 
Poland, Portugal, Spain. KAGRA is supported by Ministry of 
Education, Culture, Sports, Science and Technology (MEXT), 
Japan Society for the Promotion of Science (JSPS) in Japan; 
National Research Foundation (NRF) and Ministry of Science and 
ICT (MSIT) in Korea; Academia Sinica (AS) and National Science 
and Technology Council (NSTC) in Taiwan.

\end{acknowledgments}

\appendix

\section{Nucleon fraction inside neutron star} 
\label{sec:app-nucleon-fraction}

\begin{figure}[htbp]
\begin{centering}
\includegraphics[width=0.45 \textwidth]{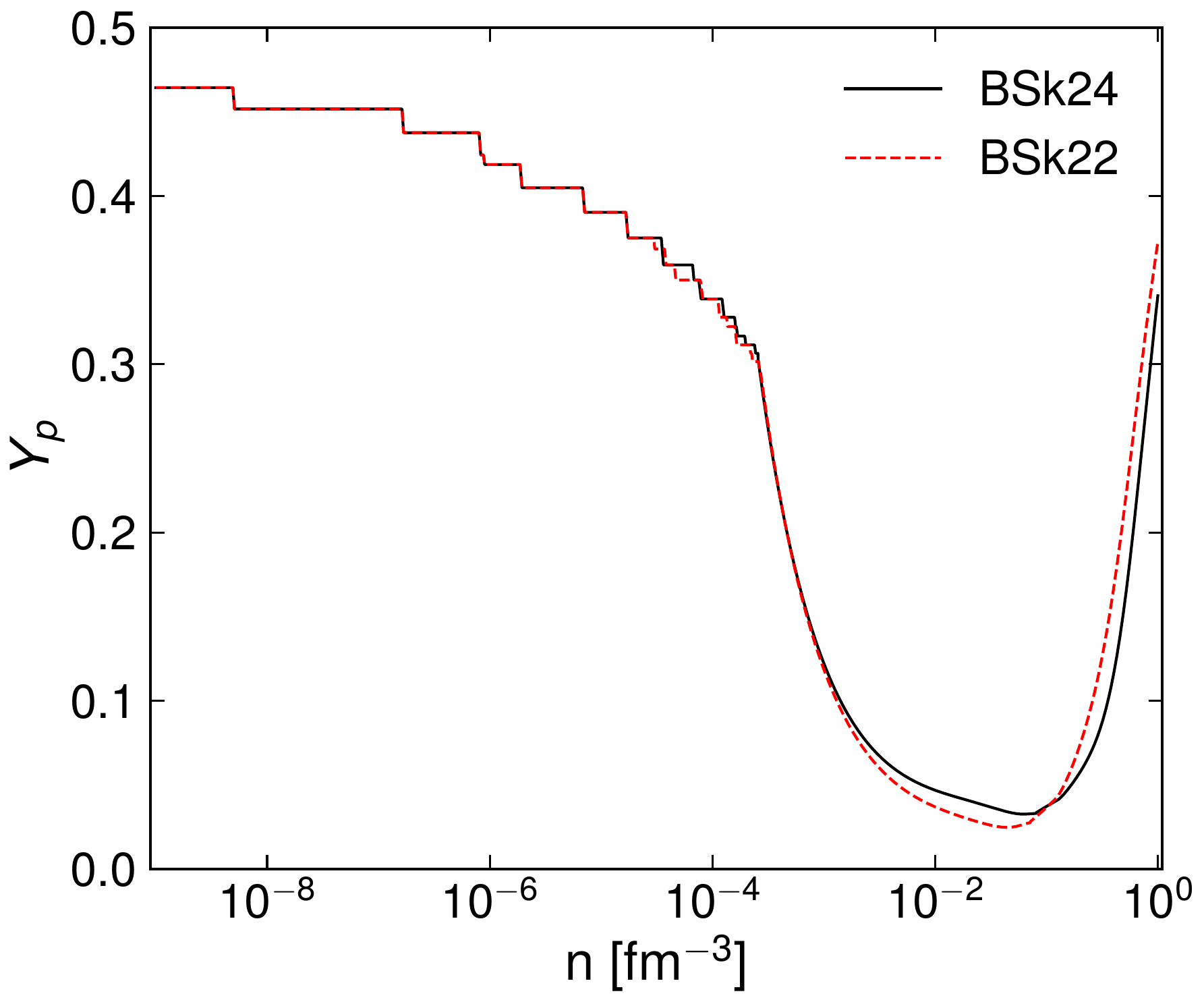}
\includegraphics[width=0.45 \textwidth]{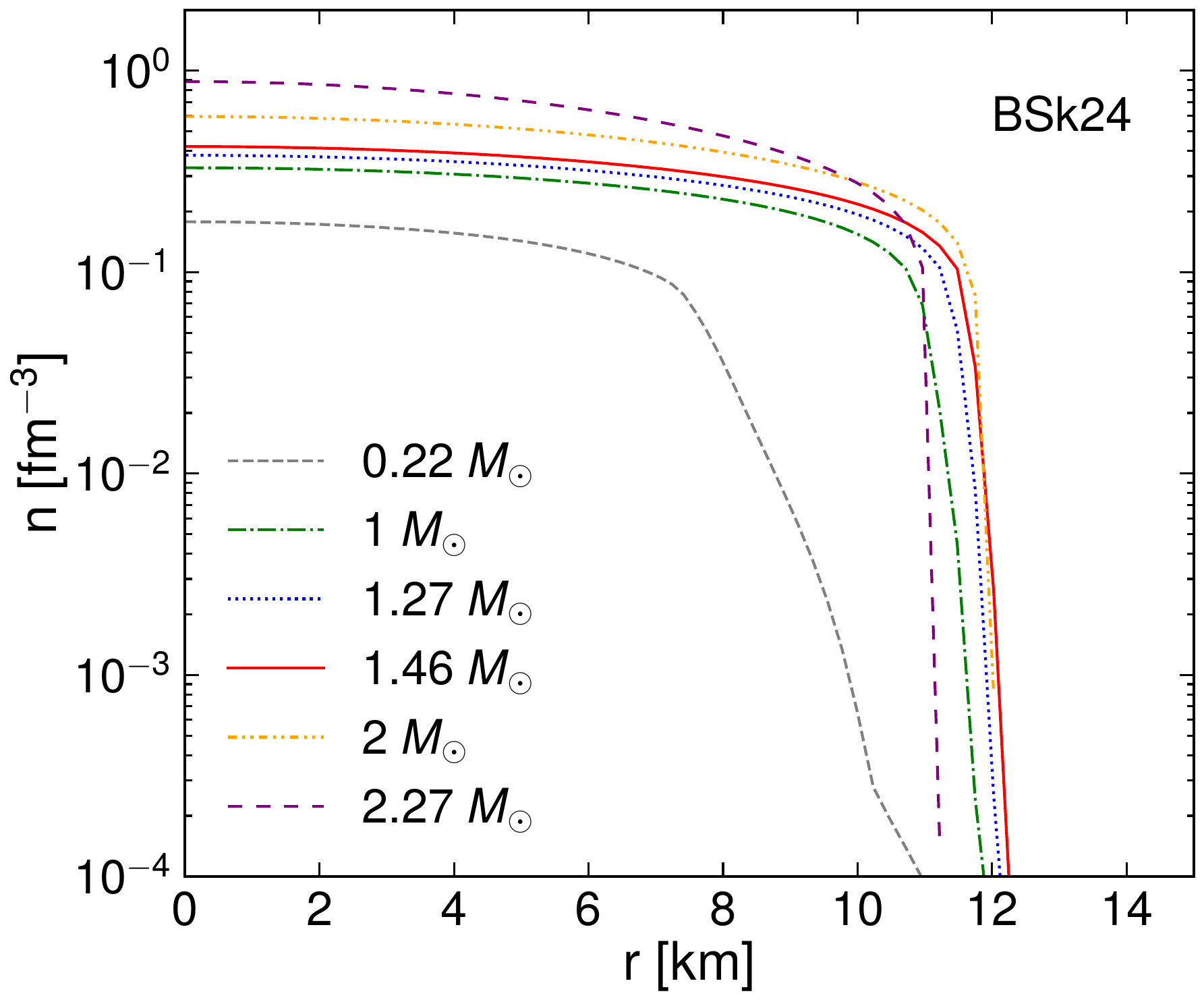}
\caption{{\bf Left:}
The proton fraction as a function of the 
nucleon number density for both BSk22
and BSk24 EoS, which are 
adopted from figure 30 of \cite{Pearson:2018tkr}.
{\bf Right:} The nucleon number density as a function of the 
radial distance for neutron stars with
different masses, 
where only the BSk24 EoS \cite{Pearson:2018tkr} 
is used.}
\label{fig:nrYpn}
\end{centering}
\end{figure}

In this section we compute the proton-to-nucleon ratio 
for the two neutron stars in GW170817. 
We compute 
the proton fraction of the NS via 
\begin{equation}\label{eq:Yp-integration}
    \left(\frac{Z}{A}\right)_{\text{NS}} = \frac{\int 4\pi r^2 Y_pn(r)dr}{\int 4\pi r^2 n(r)dr},
\end{equation}
where 
$Z$ ($A=Z+N$) is total proton (nucleon) number of the NS 
with $N$ being the neutron number,
$r$ is the radial distance, 
$n(r)$ is the nucleon number density, 
and $Y_p(n)\equiv n_p/n$ is the proton fraction 
with $n_p$ being the proton density.
The nucleon number density $n(r)$ is related to the 
energy density $\rho(r)$
via \cite{Pearson:2018tkr}
\begin{equation}\label{eq:rho-n}
    \rho(r) =n(r)(e_{\text{eq}}+m_n)
\end{equation}
where 
$e_{\text{eq}}$ is the energy per nucleon, 
and $m_n$ is the neutron mass. 
In our analysis, 
we use equation (C1) 
in Ref.~\cite{Pearson:2018tkr} to compute $e_{\text{eq}}$, 
and use the proton fraction $Y_p(n)$ in figure 30 of 
Ref.~\cite{Pearson:2018tkr}, 
which is shown in the left pangel of Fig.~\ref{fig:nrYpn} 
for completeness.

\begin{figure}[htbp]
\begin{centering}
\includegraphics[width=0.45 \textwidth]{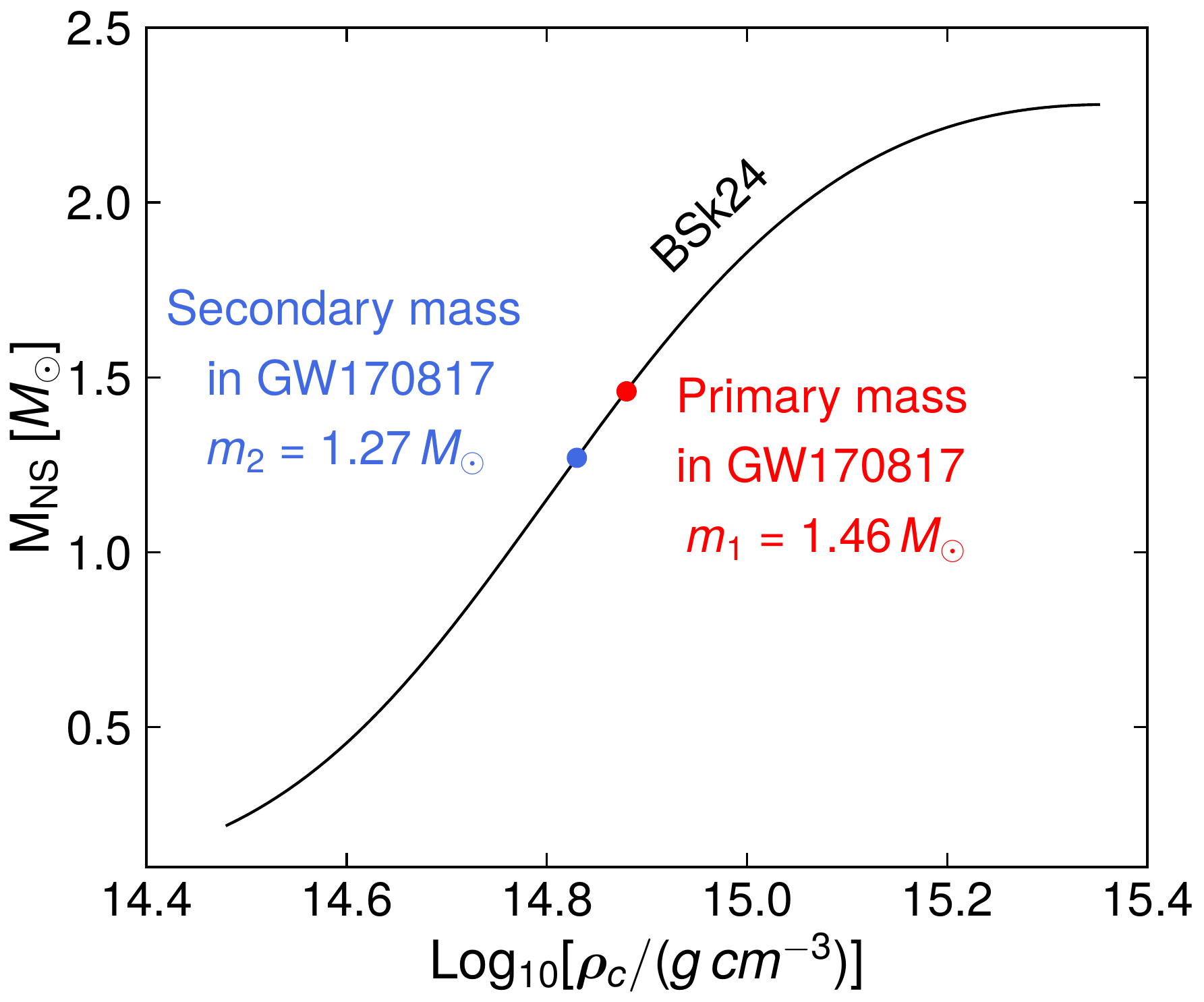}
\includegraphics[width=0.45 \textwidth]{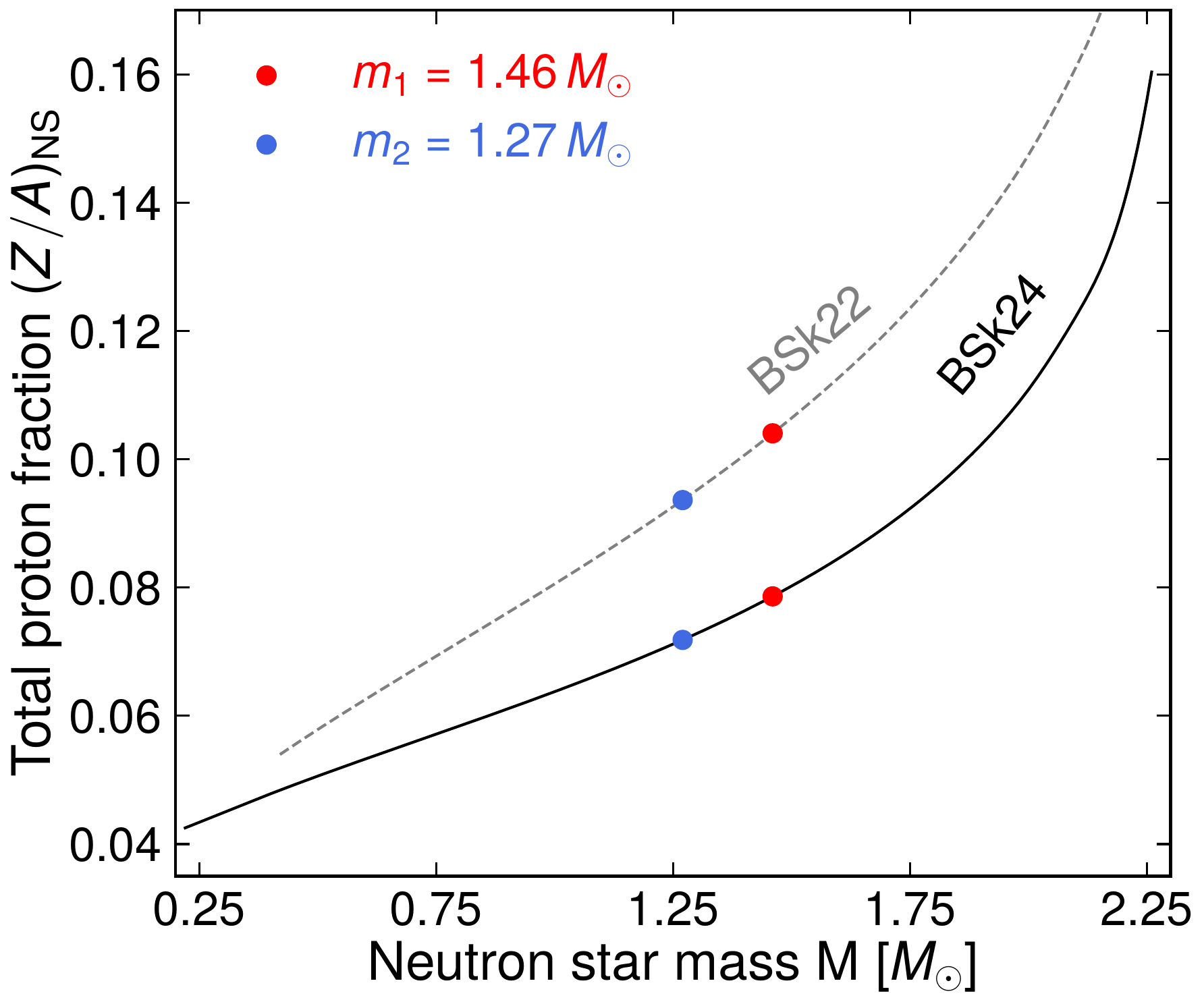}
\caption{{\bf Left:}
The NS mass as a function
of the central mass-energy density, 
calculated with the BSk24 EoS \cite{Pearson:2018tkr}. 
The primary mass $m_1=1.46\,M_\odot$
and the secondary mass $m_2=1.27\,M_\odot$ in
the GW170817 event \cite{LIGOScientific:2018mvr}
are shown.
{\bf Right:} The proton fraction 
as a function of the mass of the neutron star, 
calculated with both the BSk24 EoS (solid, black)
and BSk22 EoS (dashed, gray). 
}
\label{fig:rhocfraction}
\end{centering}
\end{figure}

To compute $\rho(r)$, we solve simultaneously 
the TOV equation 
\cite{Tolman:1939jz,Oppenheimer:1939ne}, 
the mass balance equation \cite{Haensel:2007yy}, 
and the EoS. 
The TOV equation is 
\begin{equation}
\label{eq:TOV} 
\frac{dP(r)}{dr}=-
\frac{G\rho(r) m(r)}{r^2}
\left[1+\frac{P(r)}{\rho(r)}\right]
\left[1+\frac{4\pi P(r)r^3}{m(r)}\right]
\left[1-\frac{2Gm(r)}{2r}\right]^{-1},
\end{equation}
where 
$G$ is the gravitational constant, 
$P$ is the pressure,
$r$ is the radial coordinate, 
and $m(r)$ is the mass within $r$. 
The mass balance equation for 
a spherically symmetric system is 
\begin{equation}
\label{eq:mr} 
\frac{dm(r)}{dr}=4\pi r^2\rho(r). 
\end{equation}
The EoS of a neutron star 
provides the relation between the pressure 
and the density, $P = P(\rho)$. 
For the EoS, we will primarily use the Brussels-Montreal EoS BSk24 in Ref.~\cite{Pearson:2018tkr}, 
which is considered as phenomenologically superior \cite{Pearson:2020bxz} 
among the four functionals in Ref.~\cite{Pearson:2018tkr}. 
For comparison, we also consider another EoS, BSk22. 
In our analysis we adopt the fitted $P(\rho)$ function 
given in Eq.~(C4) of Ref.~\cite{Pearson:2018tkr}.

We thus solve 
Eq.~\eqref{eq:TOV} and 
Eq.~\eqref{eq:mr}, supplemented by the EoS, 
with the boundary conditions: 
$\rho(0)=\rho_c$ and $m(\Delta r)=4\pi \Delta r^3\rho_c/3$
with $\Delta r$ being the step length. 
We use different $\rho_c$ values in 
the range of 
$\sim(10^{14.4},10^{15.4})\,\text{g}\,\text{cm}^{-3}$. 
We move from the center of the NS  
outward until $dm/dr = 0$, 
which is defined as the radius of the NS 
in our calculation. 
We thus obtain  
the NS radius $R_{\text{NS}}$, 
the NS mass $M_{\text{NS}}$, 
and the energy density profiles $\rho(r)$, 
for each $\rho_c$ value. 
We then compute the nucleon density 
$n(r)$ via Eq.~\eqref{eq:rho-n}.

The right panel figure of Fig.~\ref{fig:nrYpn} shows 
the nucleon density $n(r)$ as a function of the radial distance 
for several different NS masses. 
In additional to the 
primary mass $m_1=1.46\,M_\odot$
and the secondary mass $m_2=1.27\,M_\odot$ 
in the GW170817 event \cite{LIGOScientific:2018mvr}, 
we also provide explicit nucleon density profiles 
for neutron stars with different masses, 
which can be used to compute the proton-to-nucleon ratio 
for other possible events in the future.

The left panel figure of Fig.~\ref{fig:rhocfraction}
shows the NS mass as a function of $\rho_c$. 
We follow figure 28 of Ref.~\cite{Pearson:2018tkr} 
to show only the NS mass in the range
of $\sim (0.22,2.27)\,M_\odot$. 
Note that beyond $\sim 2.27\,M_\odot$ the BSk24 EoS becomes
hydrostatically unstable \cite{Pearson:2018tkr}.

The right panel figure of Fig.~\ref{fig:rhocfraction} shows 
the total proton fraction of the NS 
as function of the mass of the NS. 
We find that the total proton fraction 
is $Z/A \simeq 0.0786$ and $0.0718$ for 
the primary mass $m_1=1.46\,M_\odot$
and the secondary mass $m_2=1.27\,M_\odot$ 
in the GW170817 event, respectively, 
where the BSk24 EoS is used.
The total proton and neutron numbers for the 
two NSs in the GW170817 can then be obtained:  
$N_1 \simeq 1.316\times 10^{56}$, 
$Z_1 \simeq 1.536\times 10^{57}$, 
$N_2 \simeq 1.052\times 10^{56}$, and 
$Z_2 \simeq 1.359\times 10^{57}$.

We note that the NS radii computed in our program 
for $m_1=1.46\,M_\odot$
and $m_2=1.27\,M_\odot$ 
are $R_1=12.58$ km and $R_2=12.55$ km, respectively, 
which are consistent with the ones inferred 
from the observational data in the GW170817 event: 
$R_1 = 11.9^{+1.4}_{-1.4}$ km  
and 
$R_2 = 11.9^{+1.4}_{-1.4}$ km  
(at $90\%$  credible level)
\cite{LIGOScientific:2018cki}.

\section{Charge of a spherically symmetric distribution}
\label{sec:charge}

In this section we provide the derivation for Eq.~\eqref{eq:Qint}. 
From Eq.~\eqref{eq:Quniform}, one can compute the contribution 
of the shell with radius in the range of ($r$, $r+dr$), 
\begin{equation}
dQ = 
q \frac{4\pi (r+dr)^3}{3} 
\Phi \left(\frac{r+dr}{\lambda}\right)
- 
q \frac{4\pi r^3}{3} 
\Phi \left(\frac{r}{\lambda}\right) 
= q 4\pi \lambda r \sinh\left( \frac{r}{\lambda}\right) dr + {\cal O}((dr)^2), 
\end{equation}
where $q$ is the charge density such that 
for the uniform spherical 
object, one has 
$Q_{\rm pt} = q (4\pi R^3/3)$. One can then generalize the calculation 
to the case where the charge density varies with 
the radial distance $r$. 
Thus, 
the total charge of a spherically 
symmetric object with radiur $R$ is given by 
\begin{equation}
Q = 4\pi \lambda 
\int_0^R dr r q(r)   
\sinh\left( \frac{r}{\lambda}\right),  
\end{equation}
where $q(r)$ is the charge density.

\bibliography{ref.bib}{}
\bibliographystyle{JHEP}

\end{document}